# An empirical analysis of vulnerability detection tools for solidity smart contracts

Francesco Salzano[1] · Cosmo Kevin Antenucci[1] · Simone Scalabrino[1] · Giovanni Rosa[1] · Rocco Oliveto[1] · Remo Pareschi[1]

**Abstract**

The rapid adoption of blockchain technology highlighted the importance of ensuring the security of smart contracts due to their critical role in automated business logic execution on blockchain platforms. This paper provides an empirical evaluation of automated vulnerability analysis tools specifically designed for Solidity smart contracts. Leveraging the extensive SmartBugs 2.0 framework, which includes 20 analysis tools, we conducted a comprehensive assessment using an annotated dataset of 2,182 instances, manually labeled at the line level with vulnerability labels. Our evaluation highlights the detection effectiveness of these tools in detecting various types of vulnerabilities, as categorized by the DASP TOP 10 taxonomy. We evaluated the efficacy of a Large Language Model-based detection method on two popular datasets. In this case, we obtained inconsistent results with the two datasets, showing unreliable detection when analyzing real-world smart contracts. Our study identifies significant variations in the accuracy and reliability of different tools and demonstrates the advantages of combining multiple detection methods to improve vulnerability identification. We identified a set of 3 tools that, combined, achieve up to 76.78% found vulnerabilities, taking less than one minute to run, on average. This study contributes to the field by releasing the largest dataset of manually analyzed smart contracts with line-level vulnerability annotations and by conducting the largest empirical evaluation of tools to date.

## 1 Introduction

Blockchain has emerged as a disruptive technology across multiple domains since its introduction (Nakamoto 2008). A key enabler of this transformation is the adoption of Smart Contracts (SCs), which enable the automated execution of business logic on blockchains

Communicated by: David Lo.

Extended author information available on the last page of the article and are routinely used for high-stakes financial transactions. The criticality of these systems entails significant security risks, as demonstrated by well-known incidents such as the *DAO* attack, which resulted in losses of approximately $60M (Porru et al. 2017). Consequently, a substantial body of research has investigated bugs, defects, and vulnerabilities in smart contracts, producing extensive catalogs of vulnerable code patterns that may lead to unintended or catastrophic behaviors (Chen et al. 2020).

Ensuring the security of smart contracts has therefore become a central concern, motivating the development of numerous automated vulnerability detection tools. Frameworks such as SmartBugs 2.0 aggregate a large number of static and dynamic analyzers, currently including 20 tools (Di Angelo et al. 2023). However, empirical evidence has raised serious concerns about the reliability of these tools. Durieux et al. (2020), for instance, reported that 9 analysis tools flagged vulnerabilities in 97% of a corpus of 47,000 smart contracts, suggesting alarmingly high false positive rates. As a result, developers frequently resort to manual vulnerability discovery and validation, despite the availability of automated tooling (Ghaleb 2022).

The effectiveness of vulnerability detection tools is typically assessed using labeled datasets (Durieux et al. 2020; Ibba et al. 2024; Soud et al. 2023). Yet most large-scale datasets are automatically labeled based on the tools' outputs, thereby inheriting their biases and error profiles. To mitigate this issue, prior studies have released manually annotated datasets (Durieux et al. 2020; Kalra et al. 2018). Nonetheless, these efforts remain limited either in scale or in granularity. Durieux et al. (2020) provide 142 contracts annotated at the line level, while Kalra et al. (2018) supply 1,524 contract addresses labeled only at the contract level. More recently, Wang et al. (2024a) demonstrated that vulnerability detection tools achieve substantially higher effectiveness on curated benchmark contracts than on real-world smart contracts, highlighting a critical evaluation gap between controlled datasets and realistic deployment scenarios.

This gap raises a fundamental question for the community: *to what extent do existing evaluation practices overestimate the effectiveness of vulnerability detection tools when confronted with real-world smart contracts?* Addressing this question requires both realistic datasets and fine-grained ground truth annotations. To motivate our focus on granularity, we conducted a survey investigating developers' perceived utility of different levels of vulnerability annotation. The results show a clear preference for line-of-code annotations, which enable precise localization of vulnerable instructions. This distinction is crucial, as a tool may correctly identify a vulnerable function while inaccurately reporting the specific lines responsible for the vulnerability.

In this work, we address these limitations by constructing a manually annotated dataset comprising 2,182 smart contract instances, with vulnerabilities labeled at the line level. Three authors independently conducted the manual analysis, recording both the vulnerable lines and their corresponding vulnerability classes. An additional evaluator systematically double-checked contracts labeled as non-vulnerable, and discrepancies with labels from prior studies (Durieux et al. 2020; Kalra et al. 2018) were resolved through further review.

Using this dataset, we re-evaluated all 20 tools included in SmartBugs 2.0, excluding HoneyBadger, which is dedicated to honeypot detection rather than vulnerability identification. In addition to detection effectiveness, we measured each tool's execution time to assess practical usability. We further evaluated a Large Language Model (LLM)-based vulner-ability detection approach based on ChatGPT-4o, comparing its line-level performance with results previously reported by Chen et al. (2025) on the SmartBugs Curated dataset.

To assess the robustness of LLM-based detection, we explicitly considered the observation by Wang et al. (2024a) that SmartBugs Curated contracts are intentionally simple and may not reflect real-world complexity. Accordingly, we evaluated ChatGPT-4o on 400 real-world smart contracts and observed a substantial degradation in detection effectiveness compared to curated benchmarks. This result underscores the importance of evaluating vulnerability detection techniques under realistic conditions.

Finally, we investigated whether combining complementary tools can mitigate individual weaknesses. By clustering tools based on their detection behavior and selecting representative tools from each cluster, we identified a set of three tools that detect a substantially larger fraction of vulnerabilities with limited execution overhead. Throughout our analysis, we account for the arithmetic checks introduced in Solidity version 0.8.0 by considering two scenarios: one in which arithmetic vulnerabilities are counted for contracts compiled with versions before 0.8.0, and one in which they are excluded due to compiler-level protection.

Consistent with prior work (Durieux et al. 2020; Chen et al. 2025; Nguyen et al. 2023), we adopt the DASP TOP 10 taxonomy to classify vulnerabilities and evaluate tool performance.

Our empirical analysis reveals pronounced disparities in the capabilities of existing vulnerability detection tools. No single tool can detect all vulnerability classes in the DASP taxonomy. Osiris is particularly effective at identifying arithmetic vulnerabilities, while Smartcheck excels at detecting denial-of-service issues, and Solhint demonstrates strong detection capabilities for access control, unchecked low-level calls, and bad randomness, albeit with a high false-positive rate. Conkas and Slither emerge as the most effective tools under different arithmetic-check scenarios. The LLM-based approach achieves moderate recall (47.67%) on curated benchmarks, with higher precision than previously reported (Chen et al. 2025), but performs significantly worse on real-world contracts.

By clustering tools and selecting complementary detectors, we identify a combination of three tools—Conkas, Slither, and Smartcheck—that collectively detect up to 76.78% of the manually annotated vulnerabilities with limited runtime overhead.

The contributions of this work are as follows:

- We release the largest dataset of smart contracts manually annotated with line-level vulnerability labels;
- We conduct the most extensive line-level empirical evaluation to date of 19 vulnerability detection tools from SmartBugs 2.0, along with an LLM-based approach, mapped to the DASP TOP 10 taxonomy;
- We demonstrate that LLM-based vulnerability detection exhibits markedly lower effectiveness on real-world smart contracts compared to curated benchmarks;
- We identify a complementary combination of three tools that substantially improves vulnerability coverage while maintaining practical execution costs.

## 2 Background

To provide context for our study, this section reviews relevant concepts, tools, and methodologies related to SCs and their vulnerabilities.

## 2.1 Blockchain

The Blockchain is a self-governed, peer-to-peer network transaction system that enables secure operations without the need for a trusted intermediary (Alsunaidi and Alhaidari 2019). Transactions are processed on a decentralized ledger made of sequential blocks that are linked to each other, maintaining an immutable connection to their predecessors and ensuring the integrity of the chain. Each block contains validated transactions that are processed according to consensus algorithms. This ledger is shared and replicated across the network, allowing all participants to read and write data, thereby providing transparent access to the stored information for every member of the network.

## 2.2 Smart Contracts

SCs are event-driven software replicated in identical copies across decentralized nodes, designed to automatically execute code when specific conditions are met (Zou et al. 2019). The source codes of SCs are immutable. However, they can be made updateable using a proxy that routes calls to a new implementation while the original contract remains published on the blockchain, thereby preserving its immutability (Bodell et al. 2023). Once deployed, an SC is identifiable by its immutable address.

Users or other SCs can interact with SCs by invoking them through transactions. Nodes in the Blockchain network validate these transactions, and when a transaction is deemed valid, the result of executing the logic encoded in the SC is written to their local copy of the Blockchain. For inclusion in a block, nodes must execute this logic uniformly, ensuring that the stored data is irreversible due to the Blockchain's immutability. This means that if a transaction terminates unexpectedly, the outcome may not be reversible.

## 2.3 Ethereum and Gas

Ethereum is the second-largest blockchain after Bitcoin and the most widely adopted platform for SCs (Anwar et al. 2020). It supports SC execution through the Ethereum Virtual Machine (EVM), which compiles SCs written in high-level languages into Ethereum bytecode, with Solidity representing the prevalent language (Zou et al. 2019; Buterin et al. 2014), sided by Vyper.

In Ethereum, gas serves as the unit for measuring the computational effort required for transactions and interactions within the network (Zou et al. 2019). SCs are executed by miners on their nodes, who are compensated with a certain amount of gas. This compensation is provided by users initiating transactions, each of which has a *gas limit* that defines the maximum allowable gas cost. The transaction will be reverted if the gas cost exceeds this limit, triggering an exception (Chen et al. 2020).

## 2.4 Etherscan

Etherscan is a widely used blockchain explorer tailored specifically for the Ethereum network (Sasaki et al. 2024). It enables users to search, analyze, and verify numerous elements, including transactions, smart contracts, addresses, tokens, and other on-chain activities. Serving as a public ledger interface, Etherscan enhances transparency within the decentral-ized Ethereum ecosystem. It is an essential tool for developers, researchers, and security analysts who rely on it to examine and validate contract interactions, gas fees, and transaction records.

Furthermore, such a blockchain explorer provides multiple APIs that grant access to blockchain data, eliminating the necessity for users to operate their own Ethereum nodes. These APIs encompass a variety of endpoints for retrieving information related to smart contracts, such as executed transactions, source codes, bytecodes, and Application Binary Interfaces.

### 2.5 Smart Contract Vulnerabilities

In this work, we refer to SC vulnerabilities considering the DASP TOP 10[1], which is a taxonomy that encompasses the most recurrent weaknesses that we describe below.

- **Reentrancy**: Contracts can be recursively called by external contracts before state updates, causing inconsistent states.
- **Access Control**: Inadequate function authorization allows unauthorized access to private values or functions.
- **Arithmetic**: Fixed-dimension variables can overflow or underflow, compromising reliability.
- **Unchecked Low Level Calls**: Low-level calls, such as send(), call(), and delegatecall() do not propagate errors, potentially leading to undesirable outcomes.
- **Denial of Service**: Excessive gas usage can revert transactions.
- **Bad Randomness**: Predictable randomness sources in Solidity can be exploited.
- **Front Running**: Attackers can reorder transactions to their advantage.
- **Time Manipulation**: Miners can manipulate time-dependent conditions.
- **Short Address**: Shorter arguments are padded to 32 bytes, allowing data manipulation.
- **Unknowns**: Encompasses yet undiscovered vulnerabilities.

### 2.6 Smart Contract Vulnerability Detection Tools

SC analysis tools include Static and Dynamic Analysis, Formal Verification, and AI-based detection instruments (Kushwaha et al. 2022), as detailed below:

- **Static Analysis Tools**: Instruments that execute a static code examination, detecting vulnerabilities without running the contract (Feist et al. 2019).
- **Dynamic Analysis Tools**: Analysis tools and fuzzers that emulate contract interactions to reveal vulnerabilities (Nguyen et al. 2020).
- **Formal Verification Tools**: Although these tools are less frequently utilized due to their complexity, they mathematically verify the contract's properties (Murray and Anisi 2019).
- **AI-based Detection Tools**: AI-based detection tools use artificial intelligence techniques to identify vulnerabilities in SCs. These tools can learn from large datasets of contracts to detect patterns that may indicate security issues (Zhuang et al. 2021).

[1] https://dasp.co/

## 3 Related Work

This section supplies an overview of existing literature and methodologies related to vulnerability detection. To gather related work, we searched for "smart contract AND vulnerability AND detection," focusing on software engineering papers with available PDFs.

### 3.1 Foundational Studies on Smart Contract Vulnerability Detection

One of the most influential studies has been conducted by Duriex et al., who empirically evaluated 47,587 SCs with analysis tools, suggesting an elevated ratio of FP and False Negative (FN) (Durieux et al. 2020). They made such an analysis leveraging SmartBugs (Ferreira et al. 2020), at its 1.0 version, that aggregates 9 SC vulnerability analysis tools. In such a work, they considered the DAPS TOP 10 classes to establish the set of vulnerabilities. Alongside their empirical evaluation, they provided a dataset of manually annotated vulnerable SCs composed of 69 instances, while the other dataset; the one with 47K SCs, was not manually evaluated. SmartBugs received updates, including more analysis tools, since its 2.0 version, also increasing the sample of manually annotated vulnerable contracts to 142 (Di Angelo et al. 2023).

Indeed, several detection tools, based on static analysis and fuzzing, were proposed in the current literature, and some of these studies supplied labeled vulnerable SC datasets. Feist et al. presented Slither, an analysis tool dedicated to SCs vulnerability detection (Feist et al. 2019). During the evaluation, the authors compared Slither with Solhint, Securify, and SmartCheck by conducting two experiments. The first concerned two famous contracts vulnerable to reentrancy, and the second was performed on the 1,000 most used contracts. In this phase, the authors manually reviewed a random sample of 50 SCs. Tsankov et al. developed Securify, a security analyzer for Solidity SCs (Tsankov et al. 2018), to validate the tool, they manually inspected the results on a dataset composed of 100 SCs. SmartCheck is a static analysis tool proposed by Tikhomirov et al.; they evaluated it by comparing the tools with others, setting a case study based on three contracts (Tikhomirov et al. 2018). Torres et al. presented Osiris, a symbolic execution tool capable of detecting arithmetic bugs in EVM bytecode (Torres et al. 2018). To evaluate their analysis, they reused the dataset provided by Kalra et al. while proposing ZEUS (Kalra et al. 2018).

The latter team manually evaluated 1,524 contracts, considering vulnerabilities included in 5 categories of the DASP TOP 10. However, details on how they evaluated their results are missing. They built their dataset by periodically scraping *Etherscan*, *Etherchain*, and *EtherCamp*. Nevertheless, given that ZEUS translates smart contracts to the low-level intermediate language (LLVM) framework, such a work did not provide fine-grained information on the analyzed vulnerabilities, such as the location and the cause. Moreover, the tool is not available to replicate results. Replicating previous studies is a valuable way to provide information and delve deeper into specific points. In this sense, considering the sample extracted from SmartBugs Results, we replicated and extended a part of the study of Durieux et al. (2020), as they ran 9 tools (including HoneyBadger). In our study, we employed 19 tools and we compared their results with a manually evaluated ground truth, providing reliable and valuable insights into the accuracy obtained by each tool.

Li et al. evaluated 8 static analyzers for SCs to detect vulnerabilities using a set of 788 contracts, finding that vulnerabilities different from Unchecked Low Level Calls and Reen-trancy are more difficult to find by the evaluated tools (Li et al. 2024a). They evaluated the analyzers considering function-level, and a valuable fine-grained taxonomy was defined in such research. We assessed the detection ability of 11 more tools with a finer-grained granularity, releasing the widest dataset with line-level annotations.

Nguyen et al. presented MANDO-HGT, a framework dedicated to detecting SC vulnerabilities included in the DASP taxonomy (Nguyen et al. 2023). Their framework works with both the source code and bytecode of SCs and defines the contract's control flow and function-call information. This framework uses transformers to detect security issues at either the contract-level or the more fine-grained line-of-code level. They compared their line-level detection approach with six tools that we evaluated, namely, Slither, Manticore, Securify, Oyente, Mythril, and Smartcheck. Still, there is a high amount of false positives that affect the tools' results. Moreover, they considered as contracts without security concerns, all the SCs for which 11 detection tools did not find any vulnerabilities. In this scenario, our results prove that tools often fail with some class of vulnerabilities, even some improvements compared with the study of Durieux et al. (2020), which found that tools were not able to detect short addresses and bad randomness. In their work, they mentioned that they verified the labels for 423 vulnerable and 2,742 clean SCs, but details on such verification are missing. In contrast, we evaluated 19 tools (plus an LLM-based approach comparing their results on manual labels we assigned without relying on detectors' outputs that could introduce errors in the labeling phase.

## 3.2 Taxonomies and Surveys

Hejazi and Lashkari provided a broad taxonomy of 256 smart contract vulnerability detection tools, categorizing them by methodology (e.g., symbolic execution, fuzzing, machine learning) and discussing their capabilities (Hejazi and Lashkari 2025). While their evaluation focuses on a small subset of seven representative tools, our work differs by performing a fine-grained, line-level assessment on a substantially larger set of tools and vulnerabilities, enabling a more precise measurement of detection capabilities and error patterns in real-world settings.

Bresil et al. presented a comprehensive survey of deep learning models for detecting vulnerabilities in smart contracts (Bresil et al. 2025). The authors conducted a comparative and meta-analysis of various detection architectures, tools, and methodologies, assessing the impact of different components on detection rates. The study found that the field is gravitating towards the use of deep learning models, with a particular focus on hybrid neural networks (HNNs) that combine different models like CNNs, RNNs, and GNNs in series or in parallel to achieve high detection results. Such a work highlights that feature extraction of syntax and semantic information is crucial for a model's performance. the authors suggest working with opcodes because they are more readily available on the blockchain than source code.

## 3.3 New Datasets and Approaches for Vulnerability Detection

Bu et al. presented a DApp-scale approach that fine-tuned Llama3-8B and Qwen2-7B with full parameter fine-tuning and LoRA (Bu et al. 2025a), using prompts that assign auditor and verifier roles; they curate 215 DApp projects (4,998 contracts) covering only for four types of vulnerabilities, namely, reentrancy, arithmetic, timestamp dependence, and price manipulation, and mitigate imbalance with Random Over Sampling. Labels come from public security audit reports, such as Code4rena and Slowmist.

HajiHosseinKhani et al. proposed a novel approach for detecting and profiling smart contract vulnerabilities by combining an updated analyzer (Hajihosseinkhani et al. 2025), SCsVulLyzer (V2.0), with a genetic algorithm–based profiling method. Their work also introduced the BCCC-SCsVul-2024 dataset, comprising 111,897 Solidity source code samples enriched with extracted features across multiple categories to support practical validation. While the dataset is large and diverse in vulnerability types, the annotations are provided at the contract level rather than at line-level, limiting the granularity of subsequent evaluations.

Bu et al. introduced SmartBugBERT (Bu et al. 2025b), a dataset of 6,157 Ethereum SCs labeled automatically with existing tools (Oyente for reentrancy and timestamp manipulation, Maian for self-destruct, Osiris for access control), covering four vulnerability classes. On this dataset, they evaluated their method, which fuses BERT-based semantic features with bytecode-level control flow graph fragments, and achieved an F1-Score of 91.19%. However, the dataset is annotated only with the result of the tools, which are prone to FPs (Durieux et al. 2020).

LOVA (LO-cating Vulnerabilities via Attention) uses line-by-line highlighting and large language model attention to localize vulnerabilities at line-level (Li et al. 2024b), feeding attention differences into a small Bi LSTM classifier. It is evaluated on Big Vul, CVEFixes, and SmartFix. Only SmartFix includes SCs, and it contains 4 vulnerabilities of the DASP, reporting precision, recall, F1, and Top N, and it outperforms vanilla and chain of thought prompting on localization.

Recent advancements include LLM-powered vulnerability scanners dedicated to SC analysis like GptScan, an innovative tool that identifies logic vulnerabilities, which are flaws intricately related to the business logic in SCs using Large Language Models (Sun et al. 2024). Xiao et al. conducted an evaluation of 5 different LLMs, demonstrating that well-designed prompts are able to reduce the false positive rate of about 60%. Their research focused on 6 types of vulnerabilities, namely, reentrancy, DOS, access control, arithmetic, manipulated price, and oracle issues (Xiao et al. 2025). Hu et al. proposed GPTLens, a framework to enhance smart contract vulnerability detection using LLMs by performing two-stage detection based on generation and discrimination, respectively (Hu et al. 2023). In the first stage, the framework generates diverse vulnerability hypotheses, while in the second phase assesses the validity of identified vulnerabilities with the goal of reducing FPs. These authors evaluated GPTLens on 13 real-world SCs. Chen et al. used ChatGPT 3.5; 4o, and 4 to detect DASP vulnerabilities present in the SmartBugs Curated dataset. They fed the models with a prompt optimized by ChatGPT, and the results they obtained show that on such datasets ChatGPT demonstrated a high Recall with a low Precision (Chen et al. 2025). In our study, we evaluated a similar approach to evaluate ChatGPT-4o as a line-level vulnerability identifier. In this sense, we assessed the fine-grained effectiveness of ChatGPT-4o at line-level, providing insights into the variability of such effectiveness obtained by analyzing SCs from different datasets with different complexity and with or without vulnerability labels in the code hosted in public repositories.

Ibba et al. introduced a dataset of approximately 50,000 Solidity SCs, which includes both contract metrics and vulnerability data identified using the Slither static analysis tool (Ibba et al. 2024). Zheng et al. developed two datasets, namely, DAPPSCAN-SOURCE and DAPPSCAN-BYTECODE, which comprised 39,904 Solidity files and 6,665 compiled SCs, respectively (Zheng et al. 2024). A total of seven state-of-the-art SC weakness detection tools are evaluated in this study on data sources coming from 1,199 open-source audit reports. Based on such labels, they performed a file-level evaluation.

Wang et al. proposed a new tool named *ReEP* dedicated to Reentrancy detection; existing tools for detecting reentrancy vulnerabilities have high false positive rates and can improve the precision of vulnerability detection up to 83.6% (Wang et al. 2024b). To evaluate *ReEP*, the authors used two datasets, the first one comprises 34 contracts being confirmed, and the latter is the SmartBugs Curated dataset. Reentrancy attack detection has been deeply studied in research; the work of Sendner et al. is focused on that (Sendner et al. 2024). They used the SmartBugs dataset, annotating a 14k dataset, considering only reentrancy bugs using the SmartBugs dataset as a foundation. For each contract, they manually inspected the source code containing the three subtype functions (send, call, transfer). Their assessment focused on determining whether a state change occurred after the transfer of funds and whether reentrancy occurred. However, in this scenario, reentrancy is mitigated even if a modifier or a lock is used (Zhou et al. 2023b).

Finally, Wu et al. created a benchmark of 2,000 SCs from different sources to evaluate SC fuzzers (Wu et al. 2024), without formalizing the granularity of vulnerability labels, we tried to obtain this information by consulting their replication package, but it seems that the link to *terabox* with their benchmark has expired.

The reviewed studies evaluated vulnerabilities without considering a wide dataset, overcoming the function-level granularity vulnerability annotation. Thus, the lack of evaluations based on line-level posed vulnerabilities brings novelty to our study.

## 4 Motivating Study: Survey

Previous studies evaluating SC vulnerability detection tools focused on assessing detection effectiveness at the contract or function level. Findings indicate that most vulnerabilities are located in one line (Zhou et al. 2023a). For this reason, to investigate the perceived value of different levels of detection granularity, we conducted the motivating study presented in this section.

### 4.1 Survey Design

The questionnaire begins by giving the respondent a summary of its content and stating that answers can only be modified before submission. To protect privacy, we did not collect any personal data or email addresses from the respondents. It aims to gather insights from developers and researchers regarding their experiences with Smart Contracts written in Solidity, with a specific focus on identifying and fixing vulnerabilities. We conducted such a survey via Google Forms using convenience sampling, a popular method for gathering data (Bi et al. 2020; Arnaoudova et al. 2016). We reached out to researchers affiliated with other institutes, professionals employed in AstraKode[2], which is an enterprise focusing its business on blockchain, and the official Solidity Gitter. Therefore, respondents span across both the academic and industrial worlds. Survey answers and a notebook to get the survey results are included in our replication package (Salzano et al. 2024). The first three binary (Yes/No) questions explore the participants' level of experience in writing Smart Contracts, as well as identifying and resolving vulnerabilities.

[2] https://www.smau.it/partners/astrakode

The last three questions, based on 5-point Likert scales—an *ordinal* measurement scale commonly used to capture subjective perceptions (Lo et al. 2015; Yu et al. 2024), assess the respondents' opinions on the usefulness of different levels of vulnerability labeling: at the file level, function level, and line-of-code level. The scale ranges from 1 = very low to 5 = very high, reflecting increasing degrees of perceived usefulness. Table 1 lists the posed questions (Qs):

## 4.2 Survey Results

We received a total of 65 responses. Table 2 summarizes the questions regarding smart contract experience. We excluded all responses that answered "No" to the first three questions, resulting in the removal of three respondents' answers. Thus, Table 3 shows respondents' perceptions of the usefulness of labeling vulnerabilities at different levels of granularity in Smart Contracts, with most favoring fine-grained, line-level labeling (Q6), followed by function-level (Q5), and file-level (Q4) labeling. We interpret this preference as suggesting that fine-grained annotations could help developers locate and resolve issues more efficiently, which in turn motivated our study.

To further understand whether prior experience with smart contract vulnerabilities influences the perceived usefulness of different annotation levels, we grouped participants based on their answers to experience-related questions. Participants who reported having written, identified, and fixed vulnerabilities in Solidity smart contracts were labeled as Experienced (39 out of 65), while all others were grouped as Others (26 out of 39). Table 4 presents the distribution of responses across the 5-point Likert scale for file-level, function-level, and line-level annotations utility. The results highlight a general preference for finer-grained annotations, particularly at the line-level, with a larger proportion of Experienced participants rating them as highly useful. This finding supports the relevance of line-level annotation granularity in practical contexts and aligns with the motivation behind the structure of our dataset.

**Table 1** Questions of the survey

| Questions |
|---|
| Q1: Have you ever written a Smart Contract in Solidity for a software project? |
| Q2: Have you ever identified a vulnerability in your own or others' Solidity code? |
| Q3: Have you ever fixed a vulnerability in a Smart Contract written in Solidity? |
| Q4: How useful do you think it is to have a Smart Contract labeled at the file level with a given type of vulnerability? |
| Q5: How useful do you think it is to have a Smart Contract labeled at the function level with a given type of vulnerability? |
| Q6: How useful do you think it is to have a Smart Contract labeled at the line of code level with a given type of vulnerability? |

**Table 2** Responses to Solidity-related experience questions

| Question | Yes (%) | No (%) |
|---|---|---|
| Q1: Smart Contract Writing | 86.2 | 13.8 |
| Q2: Vulnerability Identification | 73.8 | 26.2 |
| Q3: Vulnerability Fixing | 66.2 | 33.8 |

**Table 3** Responses to questions on vulnerability labeling granularity considering the Likert scale

| Question | 1 (%) | 2 (%) | 3 (%) | 4 (%) | 5 (%) |
|---|---|---|---|---|---|
| Q4: Utility of Contract-level annotations | 9.7 | 25.8 | 24.2 | **30.6** | 9.7 |
| Q5: Utility of Function-level annotations | 1.6 | 1.5 | 22.6 | **53.2** | 21.0 |
| Q6: Utility of Line-level annotations | 1.6 | 0.0 | 11.3 | 19.4 | **67.7** |

**Table 4** Distribution of Likert-scale responses (1–5) for each annotation level utility, comparing participants with and without prior smart contract vulnerability experience

| Annotation Level | Experience Group | 1 | 2 | 3 | 4 | 5 |
|---|---|---|---|---|---|---|
| Contract Level | Experienced | 10.3% | 25.6% | 23.1% | **33.3%** | 7.7% |
| | Others | 7.7% | 23.1% | **30.8%** | 26.9% | 11.5% |
| Function Level | Experienced | 0.0% | 0.0% | 25.6% | **48.7%** | 25.6% |
| | Others | 3.8% | 3.8% | 15.4% | **61.5%** | 15.4% |
| Line-Level | Experienced | 0.0% | – | 5.1% | 23.1% | **71.8%** |
| | Others | 3.8% | – | 23.1% | 11.5% | **61.5%** |

# 5 Study Design

The *goal* of the study we propose is to evaluate vulnerability detection tools dedicated to analyzing Solidity SCs, comparing their results with a dataset of manually analyzed SCs extracted from datasets available in the literature. The *perspective* is that of researchers interested in identifying the most effective tools for detecting smart contract vulnerabilities. Our work is guided by the following research questions:

- **RQ$_1$:** How reliable are state-of-the-art tools in detecting vulnerabilities in Smart Contracts?
- **RQ$_2$:** To what extent does ChatGPT's effectiveness in detecting vulnerabilities differ between simple benchmark contracts and real-world cases?
- **RQ$_3$:** What is the best combination of tools to use to detect most Smart Contract vulnerabilities?

## 5.1 Study Context

The context of our study is represented by samples of the dataset introduced by Durieux et al. (2020) and Kalra et al. (2018). The former consists of about 47K SCs collected from the Ethereum Blockchain, encompassing contracts deployed since about 2018, and 142 manually annotated vulnerable contracts. In detail, each instance of the contract in this dataset has the results of each of the 9 analysis tools incorporated in SmartBugs. To the best of our knowledge, this dataset is the largest of those we found in the literature with this number of run tools against it. The dataset by Kalra et al. provides the addresses of 1,524 SCs alongside labeled vulnerability at contract-level (Kalra et al. 2018). Other works, such as the one presented by Huang et al. (2022), claimed to have greater datasets. However, the data is not publicly available. Despite our request for access to the authors, we received no response.

To classify vulnerabilities in our research, we use the DASP TOP 10 taxonomy. We motivate this choice due to the popularity of such a taxonomy, as it has been used in several studies (Durieux et al. 2020; Chen et al. 2025; Nguyen et al. 2023). In addition, this choice enables us to compare our results with those reached in the work of Durieux et al. (2020); Chen et al. (2025) since this comparison could provide insights into the accuracy of the tools. Below, we describe the procedure we used to collect Smart Contracts to consider in our dataset.

### 5.1.1 Selecting the Sources

To ensure a structured and rigorous dataset selection process, we first defined a set of criteria and then applied them to the available datasets with vulnerability annotations in the domain of SC vulnerability research. The goal was to identify all datasets that satisfied these requirements before deciding which ones to include in our study. Table 5 reports the criteria we adopted.

After identifying all publicly available datasets, which we show in Table 6, used in prior vulnerability analysis research, we systematically evaluated each of them against the criteria above.

**Table 5** Selection criteria for candidate datasets used in this study

| Criterion | Description |
|---|---|
| C1 | The dataset contains smart contracts written in the Solidity programming language. |
| C2 | The dataset contains contracts whose source code is directly or indirectly available (e.g., from repositories, materials, or Etherscan). |
| C3 | The dataset is introduced in a paper published in an A* ranked conference. |
| C4 | The paper presenting the dataset has received more than 200 citations. |

**Table 6** Overview of datasets with vulnerability annotations (Name, Reference, Criteria, Included)

| Name | Met Criteria | Included | Reference |
|---|---|---|---|
| Are We There Yet? Unraveling the State-of-the-Art Smart Contract Fuzzers | C1, C3 | ✗ | Wu et al. (2024) |
| ESCORT | C1, C3 | ✗ | Sendner et al. (2023) |
| Scrawld | C1, C2 | ✗ | Yashavant et al. (2022) |
| Smart contract vulnerability detection combined with multi-objective detection | C1, C2 | ✗ | Zhang et al. (2022) |
| Smart Sanctuary | C1, C2 | ✗ | Ibba et al. (2024) |
| SolidiFI | C1,C2,C3,C4 | ✗ | Ghaleb and Pattabiraman (2020) |
| SmartBugs Curated | C1,C2,C3,C4 | ✓ | Durieux et al. (2020) |
| SmartBugs Results | C1,C2,C3,C4 | ✓ | Durieux et al. (2020) |
| ZEUS | C1,C2,C3,C4 | ✓ | Kalra et al. (2018) |

Datasets not meeting one or more criteria were excluded. In detail, we kept the ZEUS dataset (Kalra et al. 2018), SmartBugs Curated, and SmartBugs Results datasets (Durieux et al. 2020), as the other did not match the selected criteria. The dataset used in the work of Ibba et al. (2024); Zhang et al. (2022), and *Scrawld* did not match *C3* and *C4*. Additionally, we discarded the dataset used in the evaluation of ESCORT (Sendner et al. 2023) and in the work of Wu et al. (2024), due to meeting only *C1* and *C3*. On the other hand, we decided to discard SolidiFI (Ghaleb and Pattabiraman 2020), as injected vulnerabilities are isolated from the original SC, may not accurately reflect real-world scenarios (Cai et al. 2024). To the best of our knowledge, no additional datasets met the inclusion criteria at the time of the experiments. An interesting detail that we found is that the dataset we kept comes with the vulnerabilities labels in a structured format (e.g., *.csv*, *xlsx*, or *.json*), which allows for automated analysis in a simpler way.

We ultimately retained three datasets: SmarBugs Results, SmartBugs Curated, Durieux et al. (2020), and the one by Kalra et al. (2018). As for the first one, we took into account the whole dataset of 47,587 instances on which Durieux et al. (2020) ran vulnerability detection tools to compare them. Manually analyzing all such instances would have been unfeasible. Therefore, we extracted a stratified sample, where each stratum is constituted by an SC vulnerability category of the DASP TOP 10. We acquired information about the vulnerabilities affecting each Smart Contract by relying on the output of the tools provided in this dataset. Specifically, for a given SC, we say it is (probably) affected by a vulnerability $v$ if at least two tools detected such a vulnerability for it. Notice that this step served solely to extract a stratified sample. Duplicates based on the address column are removed. Finally, stratified sampling is performed to obtain a subset of 2,737 records while maintaining the original distribution of vulnerability categories.

Hence, Table 7 summarizes the number of instances per vulnerability in the studied sample. A contract could have more than one kind of vulnerability. Moreover, according to the label assigned through this process, this sample does not contain instances vulnerable to Bad Randomness, Front Running, Short Addresses, and, for obvious reasons, Unknown vulnerabilities. This suggests that detection tools struggle to detect such vulnerabilities, as there were no instances marked as vulnerable by at least two tools, even though we found some instances during manual analysis.

One of the tools considered in SmartBugs, HoneyBadger, is not designed to find vulnerabilities; rather, it identifies honeypots — contracts that appear to have an obvious flaw but do not actually contain a vulnerability. Therefore, we considered contracts tagged as vulnerable by HoneyBadger as FPs, and thus, we have not considered them as vulnerable even if more than one tool marked them as not secure. In our sample, 6 contracts were tagged as honeypots.

**Table 7** Number of Vulnerabilities for each DASP class according to the described pre-labeling strategy in the sample extracted from SmartBugs Results

| Vulnerability | Number |
|---|---|
| Arithmetic | 2510 |
| Denial of Service | 254 |
| Reentrancy | 570 |
| Unchecked Low Level Calls | 108 |
| Time Manipulation | 87 |
| Access Control | 65 |
| Total | **3594** |

The second source we considered is the dataset provided by Kalra et al. (2018). The authors did not provide the line of code of the source code, nor the source code itself. Thus, we used the addresses contained in their dataset to obtain the source code by leveraging the *Etherscan* API, which returns the Solidity code associated with a given SC address. Then, we analyzed these SCs to tag the vulnerable line of code and merged the instances into our dataset.

We considered only the SCs having their code available on the blockchain (i.e., for which the previously mentioned API returns a valid Solidity SC). This choice ensured that we analyze the unchanged deployed code. Starting from 1,524 SCs contained in the dataset of Kalra et al., we obtained 538 valid SCs according to our selection criteria. Indeed, we excluded 981 smart contracts due to the unavailability of their source code (see C2). Some of such contracts (231) were labeled as *safe*, while the remainder 307 as *unsafe*.

Finally, the third source we considered is SmartBugs Curated[3], a labeled dataset composed of 142 manually annotated vulnerable contracts.

#### 5.1.2 Source Code Collection

The instances that compose the SmartBugs Results and Curated datasets come with the entire source code, thus we used it. Conversely, the available dataset from ZEUS provides only the address of each analyzed contract. To get the source code, we used an Etherscan API available at the following URL: https://api.etherscan.io/api?module=contract&action=getsourcecode &address=contract_address &apikey=api_key which return the source code of an SC deployed on Ethereum given a certain address. This approach ensures the reliability of the obtained data as SCs deployed on the blockchain are immutable, thus, the retrieved source code is the same as the code evaluated by Kalra et al. in their work (Kalra et al. 2018).

#### 5.1.3 Bytecode Collection

Some tools work on compiled SCs. In most cases, we have the address of the compiled SCs (e.g., for the SCs collected from the dataset by Kalra et al. and SmartBugs Results Kalra et al. 2018; Durieux et al. 2020), while in others we only have the source code (e.g., some SCs from the SmartBugs Curated Dataset). In the first case, namely when the address of the contract is available in the original datasets, in order to ensure obtaining the bytecode associated with a given source code in the most reliable way, we used two Etherscan APIs.

The first one is available at the following URL: https://api.etherscan.io/api?module=account&action=txlist &address=contract_address &startblock=0 &endblock=99999999 &sort=asc &apikey=api_key

As a response, it returns the list of transactions performed by an address, from the first block, and sorted in ascending order. For each contract, the first transaction stands as the deploy transaction, which, among several other data, contains the bytecode. To get the bytecode, we extracted the hash of the first transaction and sent it as an input to the second API,

[3] https://github.com/smartbugs/smartbugs curated

which can be accessed at the following URL: https://api.etherscan.io/api?module=proxy&action=eth_getTransactionByHash &txhash=tx_hash &apikey=api_key

This API returns the information about a transaction with a defined transaction hash, a subset of such information is illustrated in Listing 1.

```
{
  "result": [
    {
      "blockNumber": "5971135",
      "blockHash": "0xc787e7516d0d1f2699b1170475abc019500fd815313b332031df0f0f3c2d81e4",
      "timeStamp": "1531691866",
      "hash": "0x3fcf3d2f0780f729577b71417a853b021feac033ee7199b32ae5f9feb52de590",
      "nonce": "0",
      "transactionIndex": "69",
      "from": "0xf1b1747760b0a0ea0683243a44542873148b0b85",
      "to": "",
      "value": "0",
      "gas": "426978",
      "gasPrice": "7100000001",
      "input": "0x6060604052670de0b6b3a764000060015560028054600160a060020a031916730486cf65a2f2f3a392cbea398afb7f5f0b72ff4617905534156100415760080fd5b6103eb806100506000396000..."
    }
  ]
}
```

**Listing 1** A subset of the information return by the getTransactionByHash Etherscan API.

The *input* property of the result object returned in this response represents the *creator_code*, which contains the bytecode of the contract after the starting *Ox*, that we used to carry out the analysis.

Some instances of SmartBugs Curated contracts come with only the assigned contract name (i.e., FibonacciBalance.sol); to use the above-mentioned APIs, the address is mandatory. Therefore, we used the `solcx` library[4], a Python wrapper for the Solidity compiler, to generate the bytecode. The script reads the contract's source code and compiles it based on the required Solidity version, as specified in the *pragma declaration* line. This ensures that the correct compiler version guarantees compatibility and consistency in compilation results. To ensure reliability, we compared 50 solcx-compiled bytecodes with their blockchain counterparts, confirming equivalence.

The overall workflow to obtain the bytecode in both the described scenarios is depicted in Fig. 1.

## 5.2 Experimental Procedure

This section provides an overview of the methodology used to conduct the experiments.

[4] https://pypi.org/project/py-solc-x/

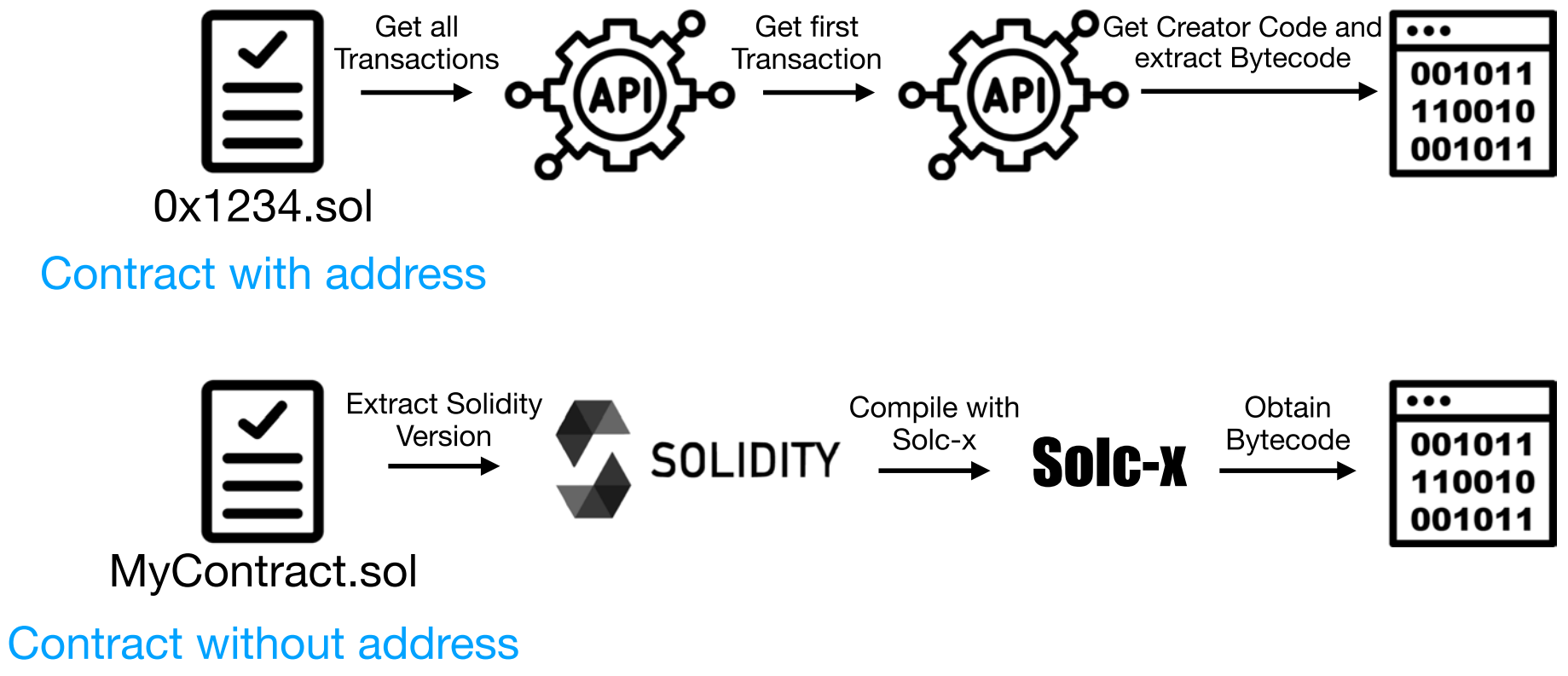

**Fig. 1** Process followed to obtain the Bytecode for contracts with and without the address

### 5.2.1 Smart Contract Vulnerabilities Tagging

To assist us in creating a high-quality dataset through manually examining SC vulnerabilities, we cloned *labeling-machine* from emadpres's GitHub[5]. The system provided in this repository allows researchers to label artifacts with minimal effort. We customized the web application to meet our specific requirements. In addition to the tag input value, we added input tags to insert the vulnerable lines, the discovered vulnerabilities, and why a selected line of code is susceptible to a specific vulnerability. The web app presents each new contract to be analyzed randomly. We stopped our analysis upon reaching 1,500 instances, deeming this number, combined with instances from other sources, sufficient for meaningful analysis, without any other significance.

We labeled vulnerabilities based on domain expertise, literature evidence (Chen et al. 2020; Zhou et al. 2023b), several vulnerable snippets reported in prior work, and the set of guidelines we previously defined in Salzano et al. (2025). To further enhance our domain expertise and deepen our understanding of Solidity security vulnerabilities, we (i) conducted a comprehensive literature review and (ii) examined vulnerable versions of smart contracts by mining fixing commits in GitHub.

(i) For the literature review, we queried Google Scholar with the string *smart contract AND fix AND (vulnerability OR defect OR recommendation)*. (ii) We mined GitHub repositories using *PyDriller* (Spadini et al. 2018), a popular framework to mine repositories, to collect commits involving security fixes. Candidate fixing commits were identified through an NLP-based filtering phase implemented with SpaCy (Explosion AI 2023), which selected commit messages referring to the resolution of security issues (e.g., fixes of reentrancy vulnerabilities). All commits returned by the filter were then manually inspected by three independent evaluators, who examined the code changes, extracted the corresponding fixing approaches, and resolved disagreements through discussion. The vulnerable versions of smart contracts preceding these fixing commits were also examined. Together with the guidelines in Salzano et al. (2025), the insights obtained from both the literature review and the mined vulnerable versions were used to define the rules and codebooks adopted to identify the labeled vulnerabilities.

[5] https://github.com/emadpres/labeling-machine

The entire vulnerability tagging process has been conducted without relying on tools; some lines were vulnerable to two vulnerabilities, for instance, to Reentrancy and Unchecked return values for low level calls. Three *validators* participated, namely, two researchers and one blockchain practitioner. After deploying the web application, evaluators identified and tagged vulnerabilities in the samples of interest. When evaluating the stratified sample extracted from SmartBugs Results, and an evaluator marked an SC as not vulnerable, a second evaluator independently validated the assessment. If both agreed, the SC was considered non-vulnerable. In cases where an evaluator labeled an SC as vulnerable, a second evaluator was assigned to review and confirm the judgment. Additionally, whenever an evaluator expressed low confidence in their labeling, the SC was double-checked by another author. In this scenario, our approach is similar to the the procedure employed by Pascarella et al. (2019) during the construction of their comment-classification dataset. In their study, one author annotated the full dataset, while a second annotator independently reviewed a subset; cases in which both annotators agreed were directly accepted.

When analyzing SmartBugs Curated and the dataset by Kalra et al., we used a different approach due to the manual evaluation already performed on the Kalra et al. and the SmartBugs Curated dataset. If an evaluator identifies a vulnerability differently from the original authors, we request a double check by another author. For instance, in the ZEUS dataset, all instances that we report as false negatives or false positives—based on a mismatch with the original labels—undergo a second validation by another author. The second author raised doubts about 16 instances, which presented a high number of vulnerabilities (12.45 on average). Nonetheless, all original labels were confirmed, and all the instances in our dataset passed consensus through a discussion among authors with conflicting opinions.

In this scenario, we agreed with the SmartBugs Curated prior manual analysis. Indeed, we confirmed all the vulnerability labels posed by a previous study (Durieux et al. 2020). Conversely, when dealing with the SCs included in the ZEUS Dataset, the point was different, because we found several vulnerabilities in contracts labeled as secure, new vulnerabilities, and some false positives. We will detail this aspect in detail in the next section.

As a result, we obtained a manually analyzed dataset containing vulnerable and non-vulnerable SCs, of which the count of vulnerable and non-vulnerable SCs is listed in Table 8. The 3 not vulnerable contracts in SmartBugs Curated refer to *Other* vulnerability category, that we discarded.

**Table 8** Number of vulnerable and non-vulnerable contracts for each source

| Source | Vulnerable contracts | Not vulnerable contracts |
|---|---|---|
| SmartBugs Curated | 140 | 3 |
| SmartBugs Results | 1083 | 418 |
| ZEUS | 311 | 227 |

#### 5.2.2 Execution of Vulnerability Detection Tools

In this section, we describe the execution of vulnerability detection tools. We used all the tools included in the popular framework called SmartBugs 2.0 for this purpose, excluding HoneyBadger due to its focus on honeypot detection. Two types of executions were performed: one on `.sol` files and the other on `.hex` files, namely, those containing the bytecode. To analyze bytecode, the `-runtime` option of the SmartBugs framework was used; this option, according to the documentation, is the one dedicated to analyzing the bytecode contained in a single line of *.hex* or *.rt.hex* files.

SmartBugs allows for the definition of the number of resources that one wants to dedicate to the tools included in the framework, for instance, the maximum RAM memory and the number of parallel processes executable. An issue that we faced regards several reboots that occurred when running Vandal, Sfuzz, and Manticore. In our preliminary analysis, we launched the tools, dedicating 12 processes to the framework, noticing reboots when such tools carried out their analysis. These issues led us to decrease the number of processes to 6, actually resolving the reboots. In addition, we assigned to the framework up to 24GB of RAM (3/4 of the available memory) by using the`-mem-limit` option. For each analysis, we set a timeout of 15 minutes per tool execution.

We focused on tools not based on AI, such as those presented by Liu et al. (2021); Zhuang et al. (2021), due to the plethora of evidence (Durieux et al. 2020; Chen et al. 2025; Li et al. 2024a) reporting the limitations of traditional detection tools and the popularity they reached. In addition, we centered our evaluation on the tools included in SmartBugs 2.0 (Di Angelo et al. 2023), which aggregates 20 well-established static and dynamic analyzers. These tools produce deterministic results, which are essential to ensure reproducibility and comparability across different studies. SmartBugs has become a widely adopted and well-established framework (Chaliasos et al. 2024; Iuliano et al. 2025), making it a suitable reference point for evaluating vulnerability detection tools.

Nonetheless, we made an exception with ChatGPT as Chen et al. presented a popular work comparing ChatGPT with some of the tools included in the set we chose for evaluation on a dataset comprised in our sample of interest. Thus, we deemed it interesting to provide insight into the accuracy of ChatGPT when using a more fine-grained level of detection and when analyzing contracts deployed in real-world, given that SmartBugs Curated instances are simple contracts with vulnerabilities, as Wang et al. (2024a) reported.

Indeed, contracts in the SmartBugs Curated dataset are intentionally designed to represent well-known vulnerabilities in a clear and isolated manner. Their simplicity lies in the linear and limited structure of the code, which facilitates the identification and understanding of the vulnerability without the complexity typical of larger or more intertwined real-world contracts. For example, Listing 2 illustrates a classic unchecked return value vulnerability in a low-level call. This makes such contracts effective for focused vulnerability detection studies but may limit their representativeness for complex real-world scenarios.

```
/*
 * @source: \url{https://smartcontractsecurity.github.io/SWC-registry/docs/SWC-104#unchecked-return-valuesol}
 * @author: -
 * @vulnerable_at_lines: 17
 */

pragma solidity 0.4.25;

contract ReturnValue {

  function callchecked(address callee) public {
    require(callee.call());
  }

  function callnotchecked(address callee) public {
     // <yes> <report> UNCHECKED_LL_CALLS
    callee.call();
  }
}
```

**Listing 2** Example of unchecked return values vulnerability (SWC-104) extracted from SmartBugs Curated.

SmartBugs tools give results by providing them as plain text. To get them in `json` format, we used the `reparse` script included in SmartBugs. Given a directory containing the file resulting from tool executions, such a script extracts structured data about the tools. When executing tools, we set the timeout to 10 minutes (15 for sfuzz, confuzzius, and manticore) as Li et al. demonstrated that tools took 5 minutes on average (Li et al. 2024a), therefore, we doubled up the time.

**Experimental Equipment and Performance Metrics** Running all the tools in SmartBugs 2.0 requires a valuable amount of time, thus, we distributed the analysis on several machines, specifically four, to speed up the analysis. However, aiming to collect uninfluenced performance in terms of time needed by each tool, we analyzed the 1,500 evaluated instances extracted from *SmartBugs Results* on a unique machine equipped with a Ryzen 9 9700x with 12 CPU cores and 24 threads and 32 GB RAM, and running Ubuntu 24.04 LTS OS. The framework configurations we reported above refer to those used on this machine.

Hence, Table 9 displays the average time taken by each tool.

To measure the reported time performance, we used the `results2csv` script included in SmartBugs, which creates a CSV file, starting from the executed tools' analysis. Such a CSV comprises several columns, among these, one column contains the duration for a certain execution that we used to obtain the average time taken by each tool. The same CSV contains an *infos* column which reported *analysis incomplete* 643 times.

### 5.2.3 Vulnerability Mapping

The analysis tools for vulnerability detection report a description of the identified vulnerabilities. To classify each description within the taxonomy proposed by the DASP TOP 10, we mapped the tool outputs to the corresponding DASP categories. Starting from the vulnerability mapping created by Durieux et al., we added 155 new mappings that were not previously provided. Tools describe the found vulnerability that can be different for the same DASP category. A mapping is a link between the description and the category. Those not yet provided are probably due to tool updates. Each new mapping has been reviewed by two authors, who assigned the class of the vulnerability independently. In the end, conflicts have been resolved through a discussion. Such a discussion involved the authors who tagged each instance that gave rise to a disagreement, until reaching a consensus between the evaluators. All conflicts were successfully resolved.

**Table 9** Detailed execution time statistics, in seconds, for security analysis tools in Solidity and Bytecode modes

| Tool | Min | Q1 | Median | Q3 | Max | Std | Mean |
|---|---|---|---|---|---|---|---|
| Solidity Mode | | | | | | | |
| Confuzzius | 0.77 | 308.37 | 592.15 | 793.34 | 1011.18 | 334.68 | 545.31 |
| Conkas | 0.63 | 1.41 | 28.03 | 62.55 | 611.07 | 95.81 | 53.19 |
| Maian | 4.60 | 55.23 | 84.36 | 145.76 | 746.69 | 103.44 | 118.56 |
| Manticore | 1.90 | 610.65 | 1010.40 | 1010.49 | 1013.93 | 393.08 | 739.69 |
| Mythril | 2.54 | 5.49 | 493.99 | 892.40 | 915.84 | 352.68 | 438.19 |
| Osiris | 0.52 | 4.50 | 43.80 | 91.76 | 1065.50 | 154.40 | 97.75 |
| Oyente | 0.62 | 3.59 | 13.43 | 21.74 | 166.06 | 18.05 | 16.79 |
| Securify | 0.45 | 26.59 | 54.24 | 96.21 | 1179.08 | 164.25 | 106.98 |
| Semgrep | 1.14 | 1.30 | 1.38 | 1.50 | 86.37 | 2.37 | 1.60 |
| Sfuzz | 0.47 | 593.17 | 596.85 | 599.60 | 1250.48 | 83.28 | 582.04 |
| Slither | 0.65 | 0.78 | 0.86 | 1.05 | 57.83 | 1.85 | 1.14 |
| Smartcheck | 1.28 | 1.45 | 1.55 | 1.73 | 147.57 | 4.07 | 1.86 |
| Solhint | 0.62 | 0.74 | 0.81 | 0.94 | 19.42 | 0.70 | 0.94 |
| Bytecode Mode | | | | | | | |
| Ethainter | 0.59 | 0.77 | 0.82 | 0.87 | 7.27 | 0.25 | 0.85 |
| Ethor | 0.62 | 4.56 | 6.72 | 10.78 | 910.94 | 52.57 | 11.41 |
| Pakala | 5.29 | 17.73 | 208.04 | 910.50 | 913.96 | 401.58 | 410.77 |
| Teether | 0.40 | 0.45 | 0.51 | 0.61 | 434.23 | 16.49 | 1.50 |
| Vandal | 0.38 | 0.53 | 0.80 | 1.26 | 1156.90 | 112.01 | 32.93 |

We report the level of agreement in terms of Cohen's Kappa, which evaluates the concordance between two assessors who categorize $N$ items into $C$ distinct categories (Cohen 1960). Conflicts have been resolved after calculating Cohen's Kappa. The acquired results reveal a very high level of agreement between the two raters, indicated by the value of Cohen's Kappa = 0.92. Notice that, the reported Cohen's Kappa value relates to mapping tool output descriptions to DASP categories, not to vulnerability label assignments on the evaluated SCs.

#### 5.2.4 RQ$_1$: Detection Evaluation

We compared the results obtained from the SmartBugs 2.0 tools with our ground truth. Specifically, for each finding from each tool, we extracted the vulnerability description and the corresponding line or set of lines of code that express it. The description was used to determine the DASP category to which the vulnerability belongs, and the line number was used to verify the match with our ground truth, if provided. Figure 2 depicts the overall workflow followed in this scenario.

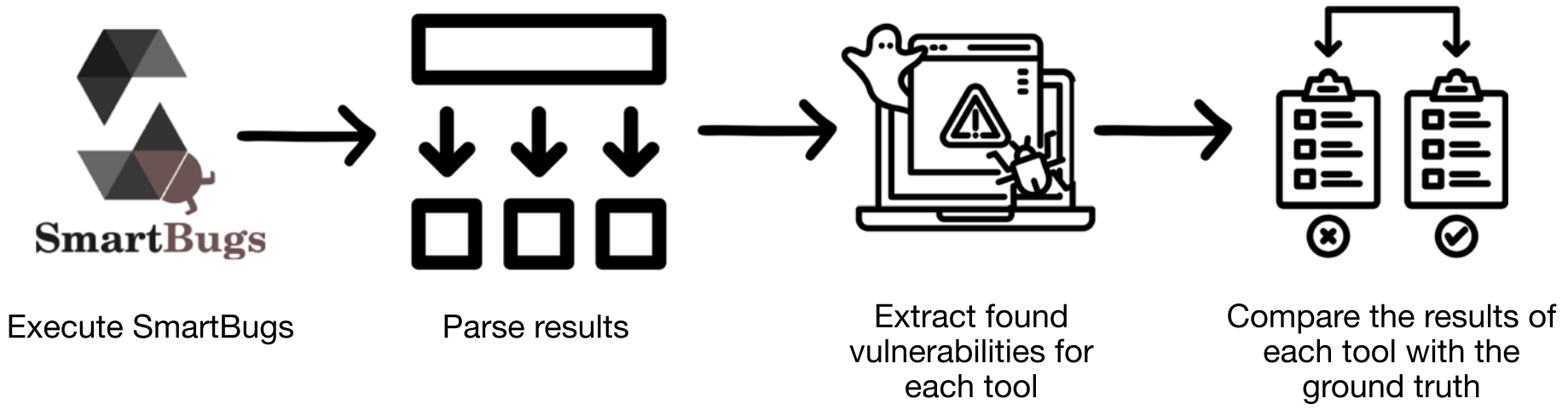

**Fig. 2** Overall workflow followed for the evaluation

We followed two different designs. If the tool's output showed the line that has the vulnerability, we considered that level of detail. Otherwise, we only considered the vulnerability itself. When delving deeper into this, bytecode analyzers point out the address of the bytecode where the tool found the weakness. On the other hand, source code analyzers highlight the specific lines of code that show the security vulnerability. Dealing with source code analyzers, after using the `reparse` script, in some cases, in the context of security and oyente executions, it did not provide the line number, to address such a miss we parsed the log file to obtain this information. Such scripts are included in our replication package (Salzano et al. 2024).

Since Solidity 0.8.0, overflow and underflow checks are performed by default, making it essential to differentiate between these two scenarios. Including arithmetic issues enables us to assess the effectiveness of detection tools in identifying problems that may still exist in older contracts. Conversely, excluding arithmetic vulnerabilities allows us to evaluate the current situation, directing our focus toward other types of vulnerabilities. To motivate our choice, we report two simple contracts that contain a function coded to perform a sum. In Listing 3 the contract uses Solidity 0.7.6, thus does not have the arithmetic default check, conversely to what occurs in Listing 4 which shows the same code, but Solidity 0.8.0 which employs the default check.

```
% SPDX-License-Identifier: MIT
pragma solidity ^0.7.6;

contract SumContract {
    // Function to sum two uint8 numbers
    function sum(uint8 a, uint8 b) public pure returns (uint8) {
        return a + b;
    }
}
```

**Listing 3** SumContract with Solidity 0.7.6.

Variables with uint8 can represent unsigned integers with 8 bits, thus a max value of 255, requesting the sum of 250 + 250 with a Solidity version less the 0.8.0, exploits the vulnerability and causes an overflow as shown in Listing 4.

```
{
  "decoded input": {
    "uint8 a": 250,
    "uint8 b": 250
  },
  "decoded output": {
    "0": "uint8: 244"
  }
}
```

**Listing 4** Transaction results of SumContract with Solidity 0.7.6, showing the arithmetic exploitation.

We report such an example with uint8 in order to represent it with small numbers to ensure readability in Listing 5.

```
// SPDX-License-Identifier: MIT
pragma solidity ^0.8.0;

contract SumContract {
    // Function to sum two uint8 numbers
    function sum(uint8 a, uint8 b) public pure returns (uint8) {
        return a + b;
    }
}
```

**Listing 5** SumContract with Solidity 0.8.0.

On the other hand, leveraging the default check introduced with Solidity 0.8.0, the same input led to a different output. In detail, the transaction execution terminates with a revert, preventing the exploitation of the arithmetic vulnerability, as shown in Listing 6.

```
{
  "decoded input": {
    "uint8 a": 250,
    "uint8 b": 250
  },
  "decoded output": {
    "0": "uint8: 0"
  },
  "revert": {
    "message": "The transaction has been reverted to the initial state
    ."
  }
}
```

**Listing 6** Transaction results of SumContract with Solidity 0.8.0, showing the arithmetic prevention.

To measure the effectiveness of the tools, we rely on accuracy, precision, recall, F1-score, and the total number of found security issues out of those we manually tagged.

### 5.2.5 $RQ_2$: Using LLM to Detect Vulnerabilities

LLMs have shown promising capabilities in several tasks related to the software code's lifecycle. For instance, in code summarization (Ahmed and Devanbu 2022), code generation (Mastropaolo et al. 2023; Corso et al. 2024), and test case generation (Dakhel et al. 2024). Chen et al. have evaluated the detection effectiveness of ChatGPT against the SmartBugs Curated dataset (Chen et al. 2025). We replicated their experiments, taking into account the results achieved by ChatGPT-4o, but we requested the model to spot vulnerability at line-level. As they did, we requested ChatGPT to optimize our prompt, setting a role and a task description followed by the input code, with the temperature set to 0. In detail, the LLM had the role of an SC security expert with the goal to find vulnerabilities and return, for each spotted vulnerability, its class and the content of the line in which it is present. Thereafter, we used such content to determine the number of each line of code declared as vulnerable. This was necessary because, during our experiments, ChatGPT appeared to struggle with counting accurately. As a result, directly requesting the number of the vulnerable line may lead to less accurate outcomes.

To respond $\mathbf{RQ}_2$, we evaluated ChatGPT-4o as an SC vulnerability detector on the SmartBugs Curated dataset and on 400 instances randomly extracted from the sample we manually analyzed coming from SmartBugs Results, a size chosen to ensure statistical significance, as 385 instances are sufficient to achieve a 95% confidence level with a 5% margin of error for large populations. We introduced this second data set in our analysis due to the findings of Wang et al., reporting that the experiments carried out obtained a lower detection rate in real-world contracts (Wang et al. 2024a).

ChatGPT-4o was selected because, at the time of the experiments, OpenAI described it as suitable for most general-purpose tasks, and it had previously been evaluated by Chen et al. (2025). We therefore assessed its performance as a general-purpose LLM, without domain-specific fine-tuning, to establish a baseline. Furthermore, through $RQ_3$, we extend the evaluation conducted by Chen et al. (2025), allowing a comparison with their results.

When evaluating ChatGPT-4o, we tried several prompts. An interesting insight is that in our first prompt, we asked ChatGPT to report the number of vulnerable lines of code. ChatGPT struggled with counting, sometimes providing numbers exceeding the total lines in the contract. We then asked ChatGPT to provide the code of the vulnerable lines, and we retrieved the numbers using a script.

To reach the prompt we used in the experimental procedure, we leveraged prompt engineering and role-based prompting. To refine the prompt, in order to limit costs, we made preliminary tries on a restricted sample of 50 SC from SmartBugs Curated. We noticed that in the system prompt, it was crucial to use role-based prompting by supplying the model with the names of categories included in the DASP TOP 10. Otherwise, ChatGPT-4o continued to label vulnerabilities with other classes, such as *Unhandled Exception*. The Chain-of-Thought strategy was also tested, but it did not mitigate this issue: even when reasoning steps were elicited, the model continued to generate labels not contained in the DASP TOP 10 taxonomy. After trying different prompts and failing to obtain satisfactory results, we set the prompt used in our experiments. Moreover, we did not rely on few-shot learning in order to avoid disaligning our experiment from the one carried by Chen et al. (2025).

Hence, the final prompt we sent to the model is reported in Listing 7, the variable code passed in the input is the source code of the SC in evaluation. To obtain it promptly, we asked the model to optimize it.

```
{
    "model": "gpt-4o",
    "temperature": 0,
    "messages": [
        {
            "role": "system",
            "content": (
                "You are an expert in smart contract security,
    specializing in identifying vulnerabilities. Use the DASP Top 10
    taxonomy: Reentrancy, Access Control, Arithmetic Issues, Unchecked
    Return Values, Denial of Service, Bad Randomness, Front Running,
    Time Manipulation, and Short Addresses."
            )
        },
        {
            "role": "user",
            "content": (
                "Analyze the following smart contract code: {code}. List
     only the line extracted by the code and the type of vulnerability
    in this format: 'foo.something(): Reentrancy; if(foo==something):
    Access Control;'. If no vulnerabilities are found, respond with 'no
    '. If uncertain between two vulnerabilities, list both. Do not
    include any extra information or use any other vulnerability classes
     outside of the DASP taxonomy. Follow the format exactly as
    instructed."
            )
        }
    ]
}
```

**Listing 7** Prompt used to ask ChatGPT-4o to find vulnerabilities.

The former prompt we used was pretty similar. The differences were related to the first idea we had, namely, to directly report the number indicating the lines of code susceptible to the found security issues. Therefore, there were differences also in the example we provided to the model to format the response, such as, for instance: `15: Access Control;` instead of `if(foo==something): Access Control;`

#### 5.2.6 RQ$_3$: Finding Optimal Tool Combination

In order to evaluate a combination of tools for effectively detecting different types of vulnerabilities in the DASP, we decided to cluster the tools based on their results. We represented the results of each tool as a vector, with each element of the vector corresponding to the vulnerabilities identified by the tool in a specific SC. The process involved transforming the reported vulnerabilities from each tool into a standardized format. For each tool, vulnerabilities were categorized using a predefined schema where each type of vulnerability was assigned a unique integer. This standardization allowed us to create a consistent numerical representation of the tool's findings.

We employed the k-means algorithm to cluster the tools' results, using the elbow method to obtain the optimal number of clusters. For each cluster, we took the best tool in terms of vulnerabilities found out of those stated by the ground truth. By virtue of this choice, we selected Conkas, Slither, and Smartcheck, one for each cluster. Clusters are displayed in Table 10, tools that provided an extremely low ratio of found vulnerabilities out of the tagged and those that presented an outcome unrelated to the DASP TOP 10 vulnerability categories are omitted from this stage. This set of excluded tools was defined on the back of the results obtained as a result of answering $\mathbf{RQ}_1$.

**Table 10** Tool classification across clusters

| Clusters | Tools |
|---|---|
| Cluster 1 | Conkas, Mythril, Osiris, Oyente |
| Cluster 2 | Confuzzius, Securify, Semgrep, Sfuzz, Smartcheck, Vandal |
| Cluster 3 | Slither, Solhint |

A combination of tools is helpful as the best-performing one does not cover all vulnerability classes, thus, K-means clustering enables the use of complementary tools based on their detection capabilities. After the analysis, Cluster 1 contains tools focused on arithmetic, reentrancy, and the return values of unchecked low-level calls. In contrast, Cluster 3 consists of tools with broader class detection capabilities. Finally, Cluster 2 comprises analyzers that provide detection results that are distinct from those of the other two clusters. Thus, the detection tool suite we used allows us to identify various DASP vulnerability categories and increase the number of vulnerabilities detected.

Some work ran several tools and got the vulnerability label according to the majority rule (Soud et al. 2023), However, this approach may not be scalable since some tools require a significant amount of time to complete an analysis. We consider the time required for the selected tools and the average time needed for each analysis. The average time we provide is calculated based on the sample extracted from SmartBugs Results, which was analyzed on a single machine.

## 5.3 Manual Analysis Outcomes and Details

We carefully manually evaluated a total of 2,182 SCs, identifying a total of 3,381 vulnerabilities. We assigned each vulnerability to the respective line of code that presents it. To the best of our knowledge, the dataset we built is the largest with this fine-grained vulnerability tagging without taking trust solely in the results of the tools. The discovered vulnerabilities are quite unbalanced, indicating that the developers may have introduced security vulnerabilities in an unbalanced manner. We fully confirmed the labels in SmartBugs Curated, but cannot quantify differences with ZEUS (contract-level) or SmartBugs Results (tool outputs). However, the obtained metrics highlight the actual differences between the tools in SmartBugs.

SmartBugs Curated includes 207 annotations, excluding 3 for "other" vulnerabilities that we discarded. Given that we confirmed all of these annotations, we manually posed 3,173 new vulnerability labels, as no other sources provided manual-assigned line-level labels. Newly tagged vulnerabilities differ from the original datasets as they are manually labeled (SmartBugs Result includes only tool outputs, and ZEUS provides only contract-level labels), so all these labels are new, resulting in significant manual work. There were 64 conflicts on instances from SmartBugs Results, 8 when analyzing ZEUS false negatives, 21 on other ZEUS instances, and none from SmartBugs Curated. We did not measure effort in terms of time, but we estimate that evaluating each contract took an average of 5 minutes.

Table 11 displays the distribution of the vulnerabilities into the DASP categories, according to the performed manual analysis. Even though several datasets and benchmarks are publicly available, the dataset we built is the largest with line-level vulnerability labels, as shown in Table 12. Moreover, as our survey highlighted that line-level vulnerability labels are perceived as more useful, we believe that the dataset we are releasing represents a valuable contribution for evaluating the effectiveness of vulnerability detectors at the level considered most relevant by academic and industrial smart contract practitioners. In addition, the dataset benefits from a manual review process. This manual curation further strengthens the quality of the ground truth, making the dataset not only the largest and the only one with line-level annotations, but also a more reliable benchmark for assessing vulnerability detection tools and LLM-based approaches.

**Table 11** Counts of manually detected vulnerabilities for each vulnerability type

| Vulnerability Class | Count |
|---|---|
| Arithmetic | 1405 |
| Reentrancy | 91 |
| Time Manipulation | 898 |
| Short Address | 1 |
| Access Control | 92 |
| Denial of Service (DOS) | 257 |
| Unchecked Low-Level Call | 583 |
| Front Running | 16 |
| Bad Randomness | 38 |

### 5.4 Finding ZEUS False Negatives Instances

ZEUS claims to be sound with zero false negatives and a low false positive rate (Kalra et al. 2018). The property of soundness in the context of program or system analysis and verification refers to the analyzer's ability to ensure that all reported results are indeed true.

However, results obtained during our manual analysis disprove this claim. Indeed, we found 49 contracts in the set of the 231 tagged as vulnerability-free in ZEUS evaluation, with vulnerabilities, thus the 21%. Among the 49 contracts we declared as false negatives of the ZEUS analysis, vulnerabilities are distributed over different categories as depicted in Table 13.

**Table 12** Overview of datasets with vulnerability annotations

| Name | Number of SC | Type of Annotations | Level of Annotations | Reference |
|---|---|---|---|---|
| BCCC-SCsVul-2024 | 111,897 | Existing source (for instance, Slither-audited SCs and SmartBugs Curated) | Contract-level | Hajihosseinkhani et al. (2025) |
| SmartBugBERT | 6,157 | Result of detection tools | Bytecode | Bu et al. (2025b) |
| SmartBugs Curated | 142 | Manual | Line-level | Durieux et al. (2020) |
| SmartBugs Results | 47,587 | Automated tool | Line-level | Durieux et al. (2020) |
| Smart Sanctuary | 50,000 | Automated tool | Interval of lines | Ibba et al. (2024) |
| Scrawld | 6780 | Automated tool | Contract-level | Yashavant et al. (2022) |
| ZEUS | 1,524 | Automated tool | Contract-level | Kalra et al. (2018) |
| SolidiFI | 1,524 | Automated tool | Injected Vulnerabilities | Ghaleb and Pattabiraman (2020) |
| Our dataset | **2,182** | **Manual** | **Line-level** | Salzano et al. (2024) |

**Table 13** Count of False Negatives values found in ZEUS dataset for each category

| Categories | Count of False Negatives |
|---|---|
| Arithmetic | 14 |
| Access Control | 1 |
| Time Manipulation | 3 |
| Unchecked Low Level Calls | 29 |

Considering the prevalence of the unchecked low-level calls vulnerability, we emphasize the importance of providing an example of vulnerable code for this weakness. In detail, the SC with the address `0xe2e4d0d3410cd3e81bfcb7dad364dd168bb499f3` presents the following function:

```
function flush() onlyowner {
        owner.send(this.balance);
    }
```

**Listing 8** False negative instance vulnerable to unchecked return value for low level calls.

In Solidity, **.send()** is a low-level call that returns a boolean indicating success or failure, so it must be checked to ensure the transaction succeeded. Failing to check the return value can lead to vulnerabilities, as errors might go unnoticed, potentially allowing exploits. Chen et al. explained this, also reporting an example that we report below, marking the line of code which exhibits the **.send()** call as vulnerable to the above mentioned vulnerability (Chen et al. 2020). Several more proofs of such a vulnerability presence in this kind of code can be found in the DASP and other research (Chen et al. 2023; Zhou et al. 2023b).

These sources also reported the fixing strategies as shown in Listing 9, highlighting one more time that the function shown above is not safe.

```
-foo.send(this.balance - 1 ether);
+if (!foo.send(this.balance - 1 ether))
+revert();
```

**Listing 9** Unchecked low level return value vulnerable and fixed example.

## 5.5 Finding ZEUS False Positives Instances

As we manually reviewed all the available instances of the ZEUS dataset, we also delved into the analysis of false positives. The authors of ZEUS acknowledged a low false positive rate, yet our findings indicate that several flagged vulnerabilities were in fact benign constructs or misclassified patterns. In total, we identified multiple contracts where the reported warnings did not correspond to any effective security issue according to the criteria established in the related literature (Chen et al. 2020; Zhou et al. 2023b).

Table 14 shows the distribution of false positives across vulnerability categories. Arithmetic and time manipulation issues emerged as the most frequently misreported, together accounting for a substantial proportion of the false positives detected during our validation process.

To illustrate the nature of these false positives, we provide in Listing 10 an example of Solidity code that was flagged as vulnerable by ZEUS, although no effective exploit or security violation is present according to manual inspection and reference guidelines.

**Table 14** Count of False Positives found in the ZEUS dataset for each category

| Categories | Count of False Positives |
|---|---|
| Arithmetic | 37 |
| Front Running | 10 |
| Reentrancy | 8 |
| Time Manipulation | 36 |
| Unchecked Low Level Calls | 3 |

```
pragma solidity ^0.4.2;

contract Store {
    address[] owners;
    mapping(address => uint) ownerBalances;

    function Store(address[] _owners) {
        owners = _owners;
    }

    function deposit() payable {
        uint ownerShare = msg.value / owners.length;
        ownerBalances[owners[0]] += msg.value % owners.length;

        for (uint i = 0; i < owners.length; i++) {
            ownerBalances[owners[i]] += ownerShare;
        }
    }

    function payout() returns (uint) {
        uint amount = ownerBalances[msg.sender];
        ownerBalances[msg.sender] = 0;

        // The send() call is checked: if it fails, funds are restored
        if (msg.sender.send(amount)) {
            return amount;
        } else {
            ownerBalances[msg.sender] = amount;
            return 0;
        }
    }
}
```

**Listing 10** Example of Solidity code reported as vulnerable by ZEUS but assessed as a false positive during manual review.

This contract was flagged by ZEUS as containing an unchecked low-level call. However, the `send()` invocation is explicitly checked through an `if` statement: if the transfer fails, the original balance is restored to avoid loss of funds. Therefore, this instance constitutes a clear example of a false positive detection.

## 6 Study Results

In this section, we report the results of our experiments and answer the three **RQs** that we posed.

### 6.1 $RQ_1$ Detection Evaluation

We underscore the results obtained to address $\mathbf{RQ}_1$, Table 15 shows the metrics resulting from our evaluation.

We distinguished the reported results considering pre and post 0.8.0 versions of Solidity, as we stated and motivated above. Tools that have no results listed in Table 15 did not detect vulnerabilities once their results were analyzed. While Manticore has not produced successful analyses, providing or segmentation faults or overcomes the timeout without terminating. Pakala, Ethainter, Ethor, and Teether concluded their analysis with nothing to report. Accuracy generally grows when not considering arithmetic vulnerabilities due to the default check. Precision has a low variation in the two scenarios. The ability of tools to detect arithmetic vulnerabilities clearly impacts their recall. Slither has shown higher recall in the scenario after the 0.8.0 update, while Conkas has the best recall when considering arithmetic weaknesses. The high number of false positives has influenced the accuracy of all tools. Mythril performed worse than the previous analysis, with several analyses reported `compilation failures of Solc` as a termination reason.

All the tools we evaluated come with a great FN count, underscoring not only that tools may fail to detect vulnerabilities but also showing the consequences of the capability of each tool to handle solely some type of security issues. Indeed, it is important to note that none of the tools included in SmartBugs 2.0 were able to detect all the vulnerabilities listed in the DASP TOP 10 taxonomy. To provide more informative and understandable results, we indicate the number of identified weaknesses per category out of the total we marked. This is done while taking into account the two scenarios based on the default arithmetic check.

Hence, Table 16 reports such results.

**Table 15** Performance metrics of the tools, empty values indicate that the tool has not detected DASP TOP 10 classes

| Tool | Accuracy | | Precision | | Recall | | F1-Score | |
|---|---|---|---|---|---|---|---|---|
| | Arithm. | No Arithm. | Arithm. | No Arithm. | Arithm. | No Arithm. | Arithm. | No Arithm. |
| *Confuzzius* | 0.1423 | 0.3241 | 0.1259 | 0.1265 | 0.1333 | 0.1495 | 0.1295 | 0.1370 |
| *Conkas* | 0.1512 | 0.1612 | 0.1419 | 0.0888 | **0.4780** | 0.2678 | 0.2189 | 0.1334 |
| Ethainter | - | - | - | - | - | - | - | - |
| Ethor | - | - | - | - | - | - | - | - |
| Madmax | 0.2952 | 0.5490 | 0.1250 | 0.125 | 0.0013 | 0.0020 | 0.0026 | 0.0040 |
| Maian | 0.2974 | **0.5501** | 0.3333 | 0.4000 | 0.0033 | 0.0040 | 0.0065 | 0.0081 |
| Manticore | - | - | - | - | - | - | - | - |
| Mythril | 0.1294 | 0.2216 | 0.0810 | 0.0638 | 0.0817 | 0.0907 | 0.0814 | 0.0749 |
| *Osiris* | 0.2156 | 0.3345 | 0.2643 | 0.1567 | 0.3714 | 0.1008 | 0.3088 | 0.1227 |
| *Oyente* | 0.0522 | 0.3481 | 0.0727 | 0.0911 | 0.0806 | 0.0513 | 0.0765 | 0.0656 |
| Pakala | - | - | - | - | - | - | - | - |
| Securify | 0.2333 | 0.4073 | 0.2649 | 0.2649 | 0.1357 | 0.2161 | 0.1794 | 0.2380 |
| *Semgrep* | 0.1026 | 0.2885 | 0.0012 | 0.0012 | 0.0005 | 0.0007 | 0.0007 | 0.0009 |
| Sfuzz | **0.3117** | 0.4800 | **0.4491** | **0.4457** | 0.3503 | 0.2860 | **0.3936** | **0.3484** |
| *Slither* | 0.1622 | 0.2054 | 0.1609 | 0.1603 | 0.4416 | **0.6817** | 0.2359 | 0.2597 |
| Smartcheck | 0.1732 | 0.3489 | 0.1790 | 0.2493 | 0.1719 | 0.2589 | 0.1754 | 0.2540 |
| Solhint | 0.0710 | 0.0974 | 0.0670 | 0.0670 | 0.2599 | 0.4219 | 0.1066 | 0.1157 |
| Teether | - | - | - | - | - | - | - | - |
| Vandal | 0.2162 | 0.3202 | 0.2848 | 0.2848 | 0.2308 | 0.3787 | 0.2549 | 0.3251 |

**Table 16** Correct detection out of manually tagged vulnerabilities of each tool per vulnerability class

| Tool | Acc. Ctrl | Arithm. | DoS | Reentr. | Unch. Call | Bad Rnd | Fr. Run. | Time Man. | Short Addr. | Found / Total | |
|---|---|---|---|---|---|---|---|---|---|---|---|
| | | | | | | | | | | Arithm. | No Arithm. |
| *confuzzius* | 1/92 | 99/1405 | 0/257 | 24/91 | 0/583 | 0/38 | 0/16 | 174/898 | 0/1 | 298/3381 | 199/1976 |
| *conkas* | 0/92 | 904/1405 | 0/257 | 47/91 | 87/583 | 0/38 | 2/16 | 254/898 | 0/1 | **1294/3381** | 390/1976 |
| madmax | 0/92 | 0/1405 | 2/257 | 0/91 | 0/583 | 0/38 | 0/16 | 0/898 | 0/1 | 2/3381 | 2/1976 |
| *maian* | 5/92 | 0/1405 | 0/257 | 0/91 | 0/583 | 0/38 | 0/16 | 0/898 | 0/1 | 4/3381 | 5/1976 |
| *mythril* | 23/92 | 57/1405 | 2/257 | 16/91 | 56/583 | 0/38 | 1/16 | 0/898 | 0/1 | 155/3381 | 98/1976 |
| *osiris* | 0/92 | **944/1405** | 0/257 | 40/91 | 0/583 | 0/38 | 3/16 | 94/898 | 0/1 | 1081/3381 | 137/1976 |
| *oyente* | 0/92 | 173/1405 | 0/257 | 28/91 | 0/583 | 0/38 | 2/16 | 41/898 | 0/1 | 244/3381 | 71/1976 |
| securify | 2/92 | 0/1405 | 0/257 | 33/91 | 245/583 | 0/38 | **8/16** | 0/898 | 0/1 | 288/3381 | 288/1976 |
| semgrep | 1/92 | 0/1405 | 0/257 | 0/91 | 0/583 | 0/38 | 0/16 | 0/898 | 0/1 | 1/3381 | 1/1976 |
| sfuzz | 3/92 | 458/1405 | 0/257 | 29/91 | 0/583 | 0/38 | 0/16 | 378/898 | 0/1 | 869/3381 | 411/1976 |
| slither | 19/92 | 80/1405 | 59/257 | 55/91 | 289/583 | 4/38 | 0/16 | **779/898** | 0/1 | 1290/3381 | **1210/1976** |
| smartcheck | 14/92 | 12/1405 | **195/257** | 44/91 | 204/583 | 0/38 | 0/16 | 1/898 | 0/1 | 470/3381 | 458/1976 |
| solhint | **33/92** | 0/1405 | 0/257 | 0/91 | 267/583 | **17/38** | 0/16 | 424/898 | 0/1 | 741/3381 | 741/1976 |
| vandal | 23/92 | 0/1405 | 0/257 | **65/91** | **538/583** | 0/38 | 0/16 | 0/898 | 0/1 | 626/3381 | 626/1976 |

As highlighted, Osiris stands as the best tool for detecting arithmetic. Nonetheless, it provides a poor ability to deal with other classes, while Smartcheck shows a particular ability to deal with Denial of Service threats. Solhint distinguished itself for revealing a high ratio of correct detection in Access Control, Unchecked Low Level Calls, and Bad Randomness. However, it comes with a higher number of FP, which makes it more difficult to make the analysis usable by SC auditors. Vandal is reliable for hunting for Reentrancy vulnerabilities and is able to find Unchecked Low Level Calls. Conkas and Slither resulted as the best tools when considering the default check of arithmetic issues, and not, respectively. The ability of Solhint to detect Bad Randomness could be a good starting point for facing Bad Randomness. Front Running is still an open problem for all tools: only Securify is able to detect any occurrence of this vulnerability, identifying 50% of the cases (8 out of 16), while all other tools detect 0%. The only Short Address vulnerability found is not enough to derive meaningful insights, and no tool is able to detect it.

Overall, the reliability of tools encompassed in SmartBugs 2.0 is highly influenced by the great number of FPs. Moreover, while some tools showed the ability to detect a considerable number of the vulnerabilities we tagged, other tools demonstrated a low ability to deal with the vulnerabilities listed in the considered taxonomy. Another important aspect is that no single tool can detect all classes of the considered vulnerabilities. This suggests that using a combination of tools, selected based on their complementary strengths, could enhance detection effectiveness.

**Summary of Findings for $RQ_1$**

Several tools, including Ethainter, Ethor, and Teether, fail to detect DASP vulnerabilities. Manticore produced no results due to analysis errors, and Pakala ended its analysis with nothing to report. Other tools like Semgrep identified indicators defined as smells and bugs, including *non-optimal-variables-swap*, which differ from DASP vulnerabilities. Conkas finds the most tagged vulnerabilities, while Slither is most effective in post-0.8+ Solidity contexts. The incidence of FPs undermines tool reliability. Additionally, FNs significantly impact effectiveness because no tool detects all categories in the taxonomy. This highlights the need for a variety of detection tools to cover different vulnerability classes. Solhint effectively addresses Bad Randomness, demonstrating improved capabilities, as the study by Duriex et al. revealed that no leading tool detected vulnerabilities in Bad Randomness and Short Addresses (Durieux et al. 2020).

## 6.2 $RQ_2$: Using LLM to Detect Vulnerabilities

We first report the results of the ChatGPT-4-based line-level detection on the SmartBugs Curated dataset, comparing them with those achieved by Chen et al. (2025). Overall, results are similar, with an average F1-Score achieved in the study of Chen et al. of 29.9%, while we achieved 31.81%. In this sample, line-level detection was more precise, but lost much recall. Hence, Table 17 details such results; notice that function-level metrics come from Chen et al. (2025).

To answer $\mathbf{RQ_2}$, we asked ChatGPT-4o to do the same task done on the contracts from SmartBugs Curated on 400 instances randomly chosen from our ground truth, the obtained results are far different. Indeed, all the considered metrics collapsed, Precision, Recall, and F1-Score reached 5.6%, 3.3%, and 1.9% respectively, as reported in Table 18.

**Table 17** Performance metrics on the SmartBugs Curated dataset for the ChatGPT-4o model on vulnerability detection, comparing Chen *et al.*'s function-level results with our line-level results (Chen et al. 2025)

| Vulnerability | Precision (%) | | Recall (%) | | F1 (%) | |
|---|---|---|---|---|---|---|
| | Function | Line | Function | Line | Function | Line |
| access control | 14.4 | 17.07↑ | 79.9 | 66.67↓ | 24.4 | 27.18↑ |
| arithmetic | 18.0 | 25.00↑ | 91.5 | 26.09↓ | 30.0 | 25.53↓ |
| bad randomness | 38.0 | 26.67↓ | 100.0 | 25.81↓ | 54.8 | 26.23↓ |
| DoS | 4.8 | 23.08↑ | 96.0 | 42.86↓ | 9.1 | 30.00↑ |
| front running | 4.9 | 0.00↓ | 70.0 | 0.00↓ | 9.1 | 0.00↓ |
| reentrancy | 30.0 | 20.86↓ | 96.6 | 87.88↓ | 45.8 | 33.72↓ |
| short addresses | 0.6 | 50.00↑ | 20.0 | 1.00↓ | 1.1 | 66.67↑ |
| time manipulation | 12.1 | 33.33↑ | 100.0 | 42.86↓ | 21.5 | 37.50↑ |
| unchecked | 59.0 | 42.42↓ | 98.1 | 36.84↓ | 73.6 | 39.44↓ |
| Average | **20.2** | **26.49**↑ | 83.6 | **47.67**↓ | 29.9 | **31.81**↑ |

**Table 18** Metrics of ChatGPT-4o model on vulnerability detection on a real-world smart contract sample we extracted from SmartBugs Results

| Vulnerability | Precision (%) | Recall (%) | F1 Score (%) |
|---|---|---|---|
| Access Control | 0.0 | 0.0 | 0.0 |
| Arithmetic | 12.0 | 1.0 | 2.0 |
| Bad Randomness | 0.0 | 0.0 | 0.0 |
| DoS | 21.0 | 1.0 | 2.0 |
| Front Running | 0.0 | 0.0 | 0.0 |
| Reentrancy | 1.0 | 4.0 | 1.0 |
| Short Addresses | 0.0 | 0.0 | 0.0 |
| Time Manipulation | 8.0 | 0.0 | 0.0 |
| Unchecked | 8.0 | 24.0 | 12.0 |
| Average | **5.6** | **3.3** | **1.9** |

As a result, we can conclude that ChatGPT has enormous differences in effectiveness while detecting vulnerabilities in simple and well-known vulnerable contracts and real-world SCs. Most SmartBugs Curated contracts are benchmarks rather than real-world deployments, limiting generalizability. Our work, however, aims to evaluate existing detection tools, not propose a new method. Such a difference in terms of detection ability has an enormous impact on the reliability of ChatGPT-based security analysis, suggesting a highly variable capability of correct detection, which depends on the analyzed contract. For this reason, we investigated the causes of this variance. First, as the complexity of each contract may influence the result of the ChatGPT analysis, we measure metrics regarding the contracts in the two samples. In their research, Chen et al. provided 7 different causes that lead to FPs (Chen et al. 2025). Analyzing those in our analysis, we add ChatGPT's tendency to present FPs in lines containing solidity feature-like code that can lead to vulnerabilities. Specifically, lines containing 'Call' for Reentrancy and Unchecked Return Values, e.g., in a function signature, are reported as vulnerable. FPs are reported even on .transfer() calls, which have been introduced to face reentrancy for the former vulnerability. This aspect further supports our decision to conduct a line-level evaluation. Since reentrancy is the most prevalent vulnerability in the SmartBugs Curated dataset, ChatGPT may be influenced by the function signature associated with the vulnerability label rather than analyzing the actual content of the function.

Hence, Table 19 shows metrics about the contract in SmartBugs Curated, while Table 20 contains the metrics related to the random sample that we extracted from SmartBugs Results. Given the high standard deviation in both samples, we report the metrics considering all the data in these samples and after removing outliers. To compute the cyclomatic complexity of the SCs in our dataset, we relied on Slither, a static analysis framework specifically designed for Solidity (Feist et al. 2019). Slither builds a complete control flow graph of each function and contract, from which it derives the cyclomatic complexity by considering the number of control flow nodes and edges.

The phase of outliers removal has been carried out by leveraging the *Interquartile Range* (IQR) method, a statistical technique used to identify and remove outliers in a dataset by relying on the quartiles ($Q_1, Q_2, Q_3, Q_4$) of the data distribution. Given a dataset sorted in ascending order, the quartiles are defined as follows:

- $Q_1$ (first quartile): the value below which 25% of the data fall.
- $Q_3$ (third quartile): the value below which 75% of the data fall.

The interquartile range is computed as:

$$IQR = Q_3 - Q_1$$

An observation $x$ is considered an outlier if it falls outside the following range:

$$Q_1 - 1.5 \times IQR \leq x \leq Q_3 + 1.5 \times IQR \tag{1}$$

where:

- $Q_1 - 1.5 \times IQR$ is the lower bound.
- $Q_3 + 1.5 \times IQR$ is the upper bound.

Any data point that lies outside this range is classified as an outlier.

As expressed in the Tables 19 and 20, there is a considerable difference in terms of the number of lines of code in the two different samples. Indeed, without removing outliers, the mean number of lines in the sample of real-world contracts coming from SmartBugs Results triples the same measure captured when dealing with the SmartBugs Curated dataset. When it comes to considering both samples passing through the phase of outlier removal, the real-world samples have twice the mean number of lines of SmartBugs Curated. A similar pattern emerges when looking at cyclomatic complexity: without removing outliers, SmartBugs Results contracts exhibit more than four times the mean complexity compared to SmartBugs Curated (65.20 vs. 15.06). Even after outlier removal, the real-world sample still maintains a substantially higher mean complexity (44.09 vs. 6.22), indicating that the real-world contracts are not only longer but also structurally more complex.

**Table 19** Statistics for the SmartBugs Curated dataset

| Metric | Value | Value |
|---|---|---|
| Metric | (Incl. Outliers) | (Excl. Outliers) |
| Total smart contracts processed | 142 | 129 |
| Mean number of lines | 102.04 | 54.02 |
| Standard deviation | 234.71 | 38.16 |
| Minimum number of lines | 14 | 14 |
| Maximum number of lines | 2470 | 183 |
| Cyclomatic Complexity max | 616 | 20 |
| Cyclomatic Complexity mean | 15.06 | 6.22 |
| Cyclomatic Complexity std. dev. | 53.81 | 4.68 |

**Table 20** Statistics for the Random Sample extracted from SmartBugs Results

| Metric | Value | Value |
|---|---|---|
| Metric | (Incl. Outliers) | (Excl. Outliers) |
| Total smart contracts processed | 400 | 355 |
| Mean number of lines | 314.14 | 191.12 |
| Standard deviation | 439.86 | 114.74 |
| Minimum number of lines | 13 | 13 |
| Maximum number of lines | 4150 | 592 |
| Cyclomatic Complexity max | 495 | 142 |
| Cyclomatic Complexity mean | 65.20 | 44.09 |
| Cyclomatic Complexity std. dev. | 77.06 | 33.76 |

To support statements about differences in complexity between the two datasets, we performed a statistical test on the cyclomatic complexity values after removing outliers using the IQR method. Since the Shapiro–Wilk test indicated that both distributions deviate significantly from normality ($p < 0.001$), we employed the Mann–Whitney U test, a non-parametric alternative to the independent samples t-test that does not assume normality and is suitable for comparing two independent groups with skewed distributions. The results show a statistically significant difference in cyclomatic complexity between SmartBugs Curated and the random sample that we used ($U = 2971.5$, $p < 0.001$), confirming that the complexity in one dataset is substantially higher than in the other. Taking account of that, we can conclude that SmartBugs Curated instances are generally smaller and simpler than the real-world contracts included in the random sample we considered in this comparison.

One more crucial aspect that may be at the root of the effectiveness variance regards the feature of SCs contained in the SmartBugs Curated dataset. Indeed, each contract contains in-code vulnerability labels as shown in Listing 11, which we removed while performing our analysis.

```
// @vulnerable_at_lines: 14
pragma solidity 0.4.25;

contract Overflow_Add {
    uint public balance = 1;

    function add(uint256 deposit) public {
        // <yes> <report> ARITHMETIC
        balance += deposit;
    }
}
```

**Listing 11** Overflow Vulnerability in Solidity

To give insight into this point, we report the min, max, and mean line counts of the contracts on which ChatGPT achieved correct detections, with and without outliers. Hence, Table 21 shows such metrics; when considering all the TPs, the mean number of lines is 67.81%, which represents the mean line count of true-positive instances expressed as a proportion of the mean line count of the entire dataset, while when cutting off outliers, the mean line count is 77.08% of the average line count observed in the full sample.

Thus, we can hypothesize that the size could influence the effectiveness of ChatGPT detection in SC vulnerability detection. To validate this hypothesis, we performed a Mann-Whitney statistical test after outlier removal, obtaining a p-value of 0.013226. Most of the data points in SmartBugs Curated were published in 2022, with some updated in April 2024. Since ChatGPT-4o's knowledge cutoff is October 2023, it is highly likely that the ones published before that date were already included in its training data. In this sense, the vulnerability annotation removal that we performed may not be enough to mitigate data leakage. To address this issue, we carried out a further analysis. In detail, we built up a set of non-real-world SCs and a set of real-world SCs. Both sets have 50 data points and have been published after the cutoff date. The former comprises vulnerable contracts from sources like Medium, while the latter is composed of real-world SCs randomly extracted from the dataset of Hajihosseinkhani et al. (2025).

While performing once again the evaluation of ChatGPT-4o on these two datasets, we obtained results that are faithful to the outcome of **RQ2**. Indeed, Table 22 shows the results related to the non-real-world SCS and real-world SCs, and the obtained metrics confirm that ChatGPT-4o is only able to deal with line-level vulnerability detection with non-real-world SCs.

**Summary of Findings for $RQ_2$**

ChatGPT has shown detection effectiveness with high variability, achieving worse results when analyzing real-world SCs. On the back of the metrics related to the size of the contracts that we collected, we can assume that such an LLM starts losing detection effectiveness when increasing the contracts' size. Moreover, using a popular dataset to evaluate LLMs may affect the evaluation results, given that models might have already seen the instances under evaluation. This could represent a future direction to be run to prove or confute this hypothesis. Furthermore, our post–cutoff analysis reinforces this finding, showing that ChatGPT-4o performs reliably only on non-real-world contracts, suggesting that data leakage and limited generalization play a key role in its effectiveness.

**Table 21** Line count statistics for all TP contracts with and without outliers

| Metric | Value (All TPs) | Value (Non-Outlier TPs) |
|---|---|---|
| Minimum number of lines | 21 | 21 |
| Mean number of lines | 213.95 | 147.32 |
| Maximum number of lines | 1020 | 379 |

**Table 22** Classification metrics for real-world and non-real-world contracts

| | Precision | Recall | Accuracy | F1 Score |
|---|---|---|---|---|
| Real-world contracts | 0.0556 | 0.1250 | 0.0400 | 0.0769 |
| Non real-world contracts | 0.6250 | 0.8163 | 0.5479 | 0.7080 |

### 6.3 $RQ_3$: Finding Optimal Tool Combination

To address $\mathbf{RQ}_3$, we emphasize the effectiveness of detection by highlighting the number of vulnerabilities identified by the top tools—in terms of the total we marked—in each cluster, specifically Conkas, Slither, and Smartcheck. Following our design, we underscore results both considering and not considering the arithmetic DASP vulnerability class. Hence, Table 23 summarizes such results. As indicated in the table, the combined operation of these tools is completed in less than one minute on average. Such a combination, based on our sample, detects 2,481 vulnerabilities out of 3,381, including arithmetic vulnerabilities, accounting for about 73%. On the other hand, in the second scenario, the used combination achieved a total of discovered vulnerabilities of 1,518 out of 1,976, representing approximately 77% of found vulnerabilities. Both these percentages overcome the effectiveness of previous tool combinations.

To statistically validate the superiority of our selected tool combination, we performed Fisher's Exact Test comparing it against each individual tool (Slither, Conkas, and Smartcheck) as well as the best combination previously reported by Durieux et al. As shown in Table 24, the results were statistically significant in all cases ($p < 0.001$), both when including and excluding arithmetic vulnerabilities. The odds ratios further confirm the strength of our combination, reaching values as high as 17.07 (vs. Smartcheck) and 13.48 (vs. Conkas) in the respective scenarios.

It is also important to emphasize that the selected tool combination is not the result of simply adding more analyzers, but arises from the clustering procedure presented earlier, which groups tools based on their complementary detection strengths. As shown in $RQ_1$, Conkas, Slither, and Smartcheck excel in distinct and non-overlapping vulnerability classes: Conkas achieves the highest recall for arithmetic vulnerabilities, Slither provides the best detection capabilities for non-arithmetic classes such as Unchecked Low-Level Calls and Time Manipulation, and Smartcheck is particularly effective for Denial of Service threats.

**Table 23** Vulnerabilities found by the best tool of each cluster and their average detection time. The total value is calculated as the union of the unique correct detections

| Tool | Total/Found | | Average Execution Time |
|---|---|---|---|
| | Arithmetic | No Arithmetic | |
| Conkas | 1294/3381 | 390/1976 | 53.19s |
| Slither | 1290/3381 | 1210/1976 | 1.14s |
| Smartcheck | 470/3381 | 458/1976 | 1.86s |
| Total | 2481/3381 **73.38%** | 1518/1976 **76.78%** | 56.19s |

**Table 24** Fisher's Exact Test results comparing the proposed combination with individual tools and the prior best combination

| Tool | With Arithmetic | | Without Arithmetic | |
|---|---|---|---|---|
| | Odds Ratio | P-value | Odds Ratio | P-value |
| Conkas | 4.45 | $5.15 \times 10^{-190}$ | 13.48 | $7.30 \times 10^{-300}$ |
| Slither | 4.47 | $2.97 \times 10^{-191}$ | 2.10 | $1.43 \times 10^{-26}$ |
| Smartcheck | 17.07 | $< 1 \times 10^{-300}$ | 10.99 | $1.51 \times 10^{-262}$ |
| Durieux et al. | 4.70 | $6.67 \times 10^{-204}$ | 5.64 | $2.69 \times 10^{-145}$ |

This complementarity is the rationale behind their selection as representatives of three different clusters.

Notably, we excluded tools such as Solhint, despite its higher recall in specific categories (e.g., Access Control and Bad Randomness), because its detection spectrum overlaps substantially with Slither when dealing with Time Manipulation and Unchecked Low Level calls and, more importantly, because it produces a much higher number of false positives, which would significantly reduce the usability of the combined results. Therefore, adding

**Summary of Findings for RQ$_3$**

Given that there is no tool encompassed in the set of state-of-the-art analysis tools evaluated that demonstrated the ability to find all the categories of security issues, using complementary tools improves the detection effectiveness. Clustering tools, considering their detection ability, enable the definition of a small set of analyzers with the capacity to outperform previous combinations proposed as a result of prior studies. Indeed, in the study of Duriex et al., the best combination found achieved the 37% of found vulnerabilities[a]. Our results, reaching up to 76%, clearly outperform these results.

[a] https://github.com/smartbugs/smartbugs-results?tab=readme-ov-file#combine-tools

## 7 Discussion and Implications

Slither demonstrated to be the most effective tool, finding the highest number of detected vulnerabilities, also trading off with the execution time. We will report the time required by all tools in our repository, which also serves as a Replication Package (Salzano et al. 2024). Generally, fuzzing tools need more time than others to terminate, while static analysis tools are shown to be quite quick.

Considering the analysis of Duriex et al., it seems that several improvements have been brought to some tools. Slither passed from detecting 22% of the vulnerabilities of their annotated dataset to 38.15% in our sample, considering arithmetic, until reaching 61.23% in the other scenario. When comparing recall in the No Arithm. configuration with the average execution time, some trade-offs become evident.

Slither achieves the highest recall (0.6817) while maintaining a very low average execution time (1.14 s), making it particularly efficient for large-scale analyses. Solhint also shows a high recall (0.4219) with a similarly low execution time (0.94 s). In contrast, Sfuzz (recall = 0.2860) and Conkas (recall = 0.2678) present similar recall values, but Sfuzz requires over 582 s on average—more than ten times longer than Conkas (53.19 s). Vandal offers competitive recall (0.3787) but at a higher execution cost (32.93 s), while Osiris and Securify achieve moderate recall (0.1008 and 0.2161, respectively) with execution times around 97.75 s and 106.98 s.

These results highlight that tool selection should account not only for detection capability but also for computational efficiency, especially when processing large datasets or under strict time constraints. None of the analysis tools used can detect all of the vulnerabilities of the DASP. Additionally, some tools, like Manticore and Sfuzz, require a considerable amount of time for each analysis, with the former not providing highly accurate results. Bad Randomness, Short Address, and Front Running attacks are still hardly detected, even though our results show that some tools are able to detect Bad Randomness, and Securify is able to find Front Running, contrary to what is demonstrated by Durieux et al. (2020). This comes as an outcome of the tools' update and new tools included in SmartBugs, tracing the way for more comprehensive vulnerability analysis.

Correct detection often depends on contextual analysis. The adopted methodology explicitly accounts for this challenge: tools such as Mythril and Slither analyze the surrounding execution context and can distinguish external calls followed by state-changing operations, thereby enabling the identification of patterns that may lead to reentrancy attacks.

The used LLM has demonstrated high variability in detection effectiveness with the two analyzed samples. There may be several motivations causing this. First, contracts in SmartBugs Curated were often used as ground truth; thus, ChatGPT could have learned where vulnerabilities are. Moreover, such contracts have information in source code underscoring the vulnerability location and category that we removed in our analysis, as Chen et al. have done Chen et al. (2025). One more aspect to consider is that SmartBugs Curated vulnerabilities may be more straightforward than those in real-world scenarios, especially for LLMs as some instances are replicated in other popular repositories used as a benchmark for security and analysis tools like *not-so-smart-contracts*[6].

What we cannot control is the possibility of LLMs having been trained on such public data, learning to deal with these contracts guided by the security annotation provided in their code. This paves the way for future studies dedicated to comparing the effectiveness of LLM in highlighting vulnerabilities between contracts encompassed in public datasets, on which models may have performed the training state, and contracts with similar complexity but out of the training data due to privacy or publication after the learning cut-off date.

SCs are typically deployed on public blockchains, meaning their source code is often publicly accessible. This availability provides researchers with an enormous dataset of real-world SC code to analyze. AI-based vulnerability detection has the potential to address one of the major challenges in this field: the high number of false positives produced by traditional approaches. However, the effectiveness of such AI-based methods depends heavily on the availability of large, high-quality labeled datasets. A substantial amount of accurately annotated data is therefore essential to ensure reliable training and evaluation of detection models.

The dataset we are releasing may appear to be slightly larger than the one published by Kalra et al.; when considering this point, it should be noticed that we provide the annotation with a line-of-code level labeling, instead of the label at the level of the entire contract. On the other hand, there are published studies on larger samples, but none of these used line-level labeling posed by manual revision. Our more granular vulnerability tagging obtains more relevance since, as shown by Zhou et al., vulnerability can be patched with one-line modification (Zhou et al. 2023b), and on the basis of the results of the motivating study we conducted.

**Implications For Practitioners** Select tools by balancing recall and throughput. For large-scale screening, start with a fast, high-recall tool (e.g., Slither), and then apply a slower, precision-oriented analyzer to the flagged lines. Where tools show similar recall but very different costs (e.g., Conkas vs. Sfuzz), prefer the cheaper option in CI and reserve expensive analyses for audits.

[6] https://github.com/crytic/not-so-smart-contracts/tree/master

**Implications For Tool Builders** Line-level outputs materially aided by pointing to concrete locations. Emitting the line content reduces parsing or counting errors downstream. Support easy composition of fast, high-recall pre-scans with precise verifiers, and report execution time alongside findings to enable cost-aware deployment.

**Implications For Dataset Curators and Benchmarks** Underrepresented classes (e.g., Bad Randomness with 38 instances; Front Running with 16) limit generalizability. Future benchmarks should balance classes or, at minimum, report per-class metrics and macro-averages. Providing line-level ground truth may be crucial to evaluate localization quality and to train models beyond coarse contract-level labels.

**Implications for Researchers** General-purpose LLMs are a realistic baseline, but their precision–recall profile and sensitivity to prompts require replication on real-world code; thus, releasing per-instance outcomes (TP, FP, TN, FN) and line-level annotations would enable stronger analyses.

# 8 Threats to Validity

## 8.1 Construct Validity

Construct validity threats arise from potential errors in vulnerability labeling. To address this, two evaluators independently tagged each instance. We did not rely on tools during labeling to avoid affecting our analysis with false positives, except when we used their output to select a stratified sample, which was a proxy, but we mitigated the impact by analyzing a relevant sample manually. Our stratified sample excluded Bad Randomness (38 cases) and Front Running (16 cases) due to their low frequency. This may affect the generalizability of the results, especially for tools targeting these specific vulnerabilities.

## 8.2 Internal Validity

A possible subjectivity may have been introduced in the manual analysis, which was mitigated by using multiple evaluators. We obtained the code of some contracts via the address, which may create errors; to mitigate this, the data were obtained by querying the blockchain, which is immutable. To verify the correct functioning, we used this strategy on some contracts we obtained from Ethereum, and found equivalences.

It is important to note that most SmartBugs Curated contracts are not deployed on the blockchain and therefore do not represent real-world smart contracts. Those that are deployed are inactive. Instead, these contracts are intended to serve as a benchmark for evaluating the effectiveness of vulnerability detection tools. While this raises concerns about the generalizability of LLM-based detection, our study does not propose a new approach but rather focuses on conducting an evaluation. The simplicity and synthetic nature of the SmartBugs Curated contracts may limit the external validity of our findings, as they do not fully represent the complexity of real-world smart contracts.

When re-evaluating the manually labeled contracts from Kalra et al. (2018), a potential threat is that a false negative may remain unnoticed if both the original authors and our first evaluator independently fail to identify the vulnerability. This risk applies only to contracts that Kalra et al. labeled as non-vulnerable. Although this scenario cannot be entirely excluded, it requires two independent misjudgments, which we consider unlikely given the prior manual analysis and our additional review step.

### 8.3 External Validity

The sample under study may not fully reflect the actual situation; therefore, to mitigate this risk, we selected a statistically relevant sample. Moreover, we relied on literature-known sources that have been previously used to evaluate tools. Regarding the LLM-based evaluation, ChatGPT may provide different answers to equivalent requests; therefore, full replicability of this part of the study is subject to inherent limitations. Finally, regarding our survey, we conducted it using convenience sampling, which might introduce bias; however, this method is widely accepted and practical for gathering data (Arnaoudova et al. 2016; Bi et al. 2020).

## 9 Conclusion

In this paper, we provide a dataset of 2,182 Solidity SCs that are manually reviewed to tag vulnerabilities on a specific line of code to meet the results of the survey we crafted. Analyzed SCs have been extracted from state-of-the-art datasets. We present the largest dataset with line-level vulnerability tags on which we evaluated 19 state-of-the-art tools; even in this case, our study is the one that evaluated the greatest number of tools against a manually annotated sample.

Moreover, we report performance metrics in terms of time taken for each tool per execution on average. Additionally, our work provides a combination of 3 tools to improve the number of detected vulnerabilities, also keeping the analysis scalable, reaching the percentage of 76.78% of detected security concerns.

As Software Engineering is walking in the LLM era, we measured the effectiveness of using ChatGPT when receiving requests to find vulnerable lines of code in Solidity SC. We evaluate it with instances coming from two popular datasets, namely SmartBugs Curated and SmartBugs Results. When spotting vulnerabilities in the first dataset, ChatGPT achieved satisfactory results; the lower Recall obtained with respect to the study of Chen et al. is likely due to the granularity level that we used. Surprisingly, Precision was slightly better. Conversely, the analysis of instances extracted from the latter dataset showed poor capability detection with real-world SCs, reporting a low reliability in this task.

Previous empirical evaluation of SC analysis tools reported that no tools detected Bad Randomness, Front Running, and Short Addresses vulnerabilities (Durieux et al. 2020). Despite such an improvement, the correct finding of such weakness classes has to be improved.

Future directions should explore strategies to improve the number of correctly detected vulnerabilities. The high rate of FPs is still an open problem, even if ReEP achieved enhanced Precision (Wang et al. 2024b), this contribution is limited to Reentrancy, and solutions other than the majority rule should be explored to obtain a good trade-off between Recall and Precision. Precision-oriented techniques developed for Reentrancy should be extended to other categories. Moreover, better balanced benchmarks for underrepresented classes and the release of per-instance outcomes (TP, FP, TN, FN) would enable sound statistical testing.

Worse detected vulnerabilities are related not only to code but also to Blockchain mechanisms, thus depending not only on code but also on the aspects associated with the miner's activities. Integrating transaction-level simulation that models mempool dynamics, transaction ordering, validator policies, timestamp drift, and gas constraints, together with replay on forked chains or sandboxes with realistic state, can help validate alerts that depend on environment and transaction ordering assumptions.

**Author Contributions** Conceptualization: Francesco Salzano, Simone Scalabrino; Methodology: Francesco Salzano, Simone Scalabrino; Formal analysis and investigation: Francesco Salzano, Cosmo Kevin Antenucci, Giovanni Rosa; Writing - original draft preparation: Francesco Salzano; Writing - review and editing: All authors; Funding acquisition: Simone Scalabrino, Rocco Oliveto, Remo Pareschi; Resources: Simone Scalabrino, Rocco Oliveto, Remo Pareschi; Supervision: Simone Scalabrino, Rocco Oliveto, Remo Pareschi.

**Funding** Open access funding provided by Università degli Studi del Molise within the CRUI-CARE Agreement. This work is funded by PRIN Project Trust Machines for TrustlessNess (TruMaN): The Impact of Distributed Trust on the Configuration of Blockchain Ecosystems (Identifier Code 2022F5CLN2– CUP H53D23002400006) financed by the Italian Ministry of University and Research and by the National Recovery and Resilience Plan (NRRP) and by the Italian Ministry of University and Research, project SOP (Securing sOftware Platforms - CUP: H73C22000890001), as part of the SERICS project (Security and Rights in CyberSpace - n. PE00000014 - CUP: B43C22000750006).

**Data Availability Statements** The datasets generated and analyzed during the current study are available in the replication package of the study, available at: https://github.com/fsalzano/Empirical-Analysis-of-Vulnerability-Detection-Tools-for-Solidity-Smart-Contracts.

## Declarations

**Conflict of Interests** The authors declare no conflict of interest.

**Ethical Approval** Not applicable.

**Informed Consent** We conducted a survey involving human participants to gather their opinions on smart contract vulnerability labeling. Participation was entirely voluntary, and no personally identifiable information was collected. The survey clearly stated that, by clicking the "Submit" button, respondents agreed to the treatment of their responses for research purposes. This approach to informed consent was deemed sufficient given the non-invasive nature of the survey and the anonymity of the collected data.

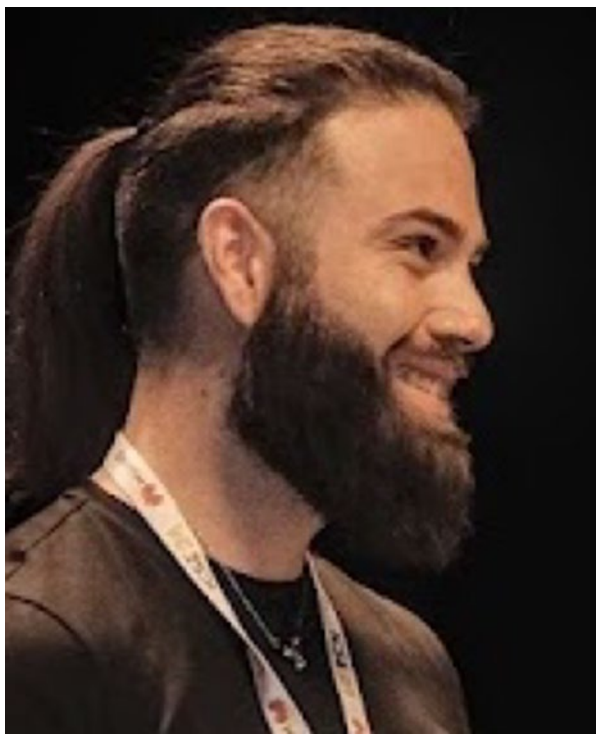

**Francesco Salzano** is an Information Engineer and a Ph.D. Candidate at the University of Molise (Italy). His research focuses on blockchain-based system architecture, software engineering for blockchain, and smart contracts. He mainly dedicates his research effort to vulnerability detection and mitigation in smart contracts, with a particular emphasis on open-source developer practices. Additionally, he explores the application of Artificial Intelligence and Large Language Models to enhance security and clone detection for smart contracts deployed on the blockchain. His professional activity involves several collaborations with IT companies as an IT architect, project manager, and full-stack developer.

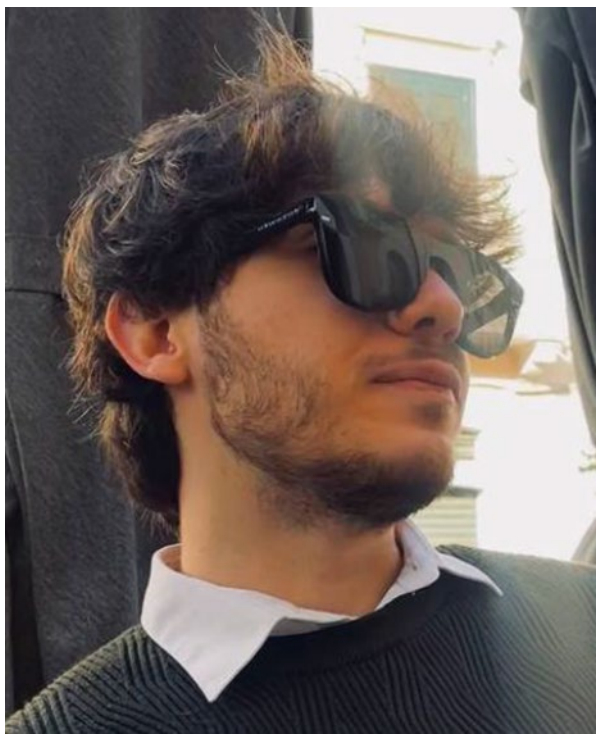

**Cosmo Kevin Antenucci** received his Bachelor's degree in Computer Science from the University of Molise, where he developed a thesis on databases under the supervision of Prof. Remo Pareschi and Dr. Francesco Salzano. He is currently a Software Developer specializing in embedded systems.

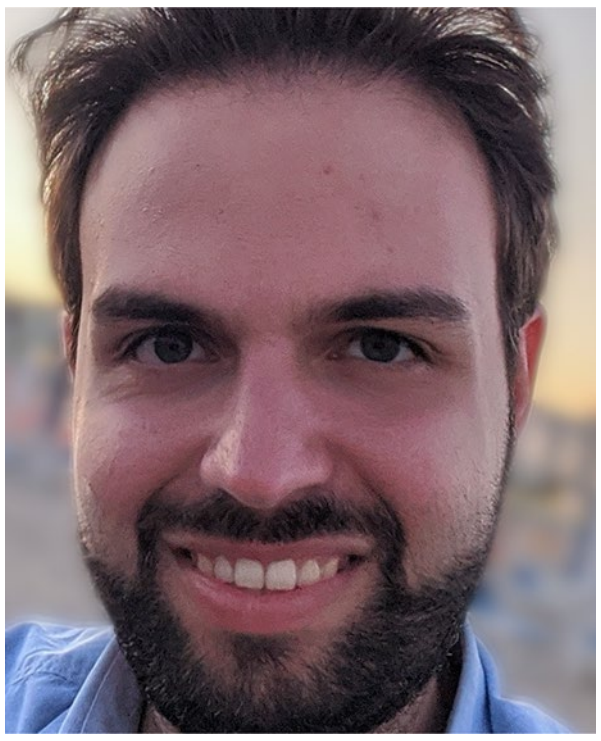

**Simone Scalabrino** received his PhD from the University of Molise (Italy) in 2019, defending a thesis on automatically assessing and improving source code readability and understandability. He is an Assistant Professor at the University of Molise, where he is the founder and director of the DEVeloper-centrIc Software Engineering Research group (DEVISER). His main research interests include code quality, software testing, and empirical software engineering. He has received three ACM SIGSOFT Distinguished Paper awards at ICPC 2016, ASE 2017, and MSR 2019. He regularly contributes to the international software engineering community as a member of various conference organizing and program committees.

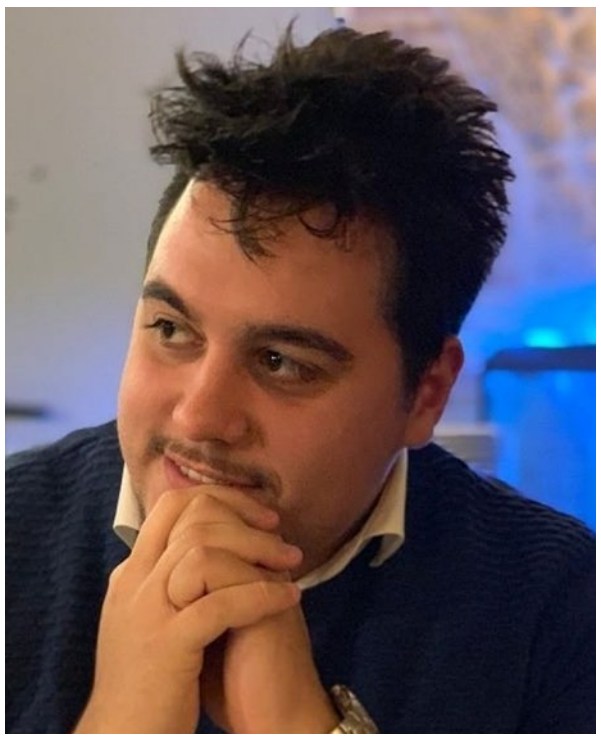

**Giovanni Rosa** is a Postdoctoral Researcher in AI for Software Engineering at Universidad Rey Juan Carlos, Spain. He received his Ph.D. in Software Engineering from the University of Molise, Italy, in April 2024. His research activity focuses on AI for Software Engineering, Mining of Software Repositories, software quality and maintenance, and Empirical Software Engineering. More information is available at https://giovannirosa.com/.

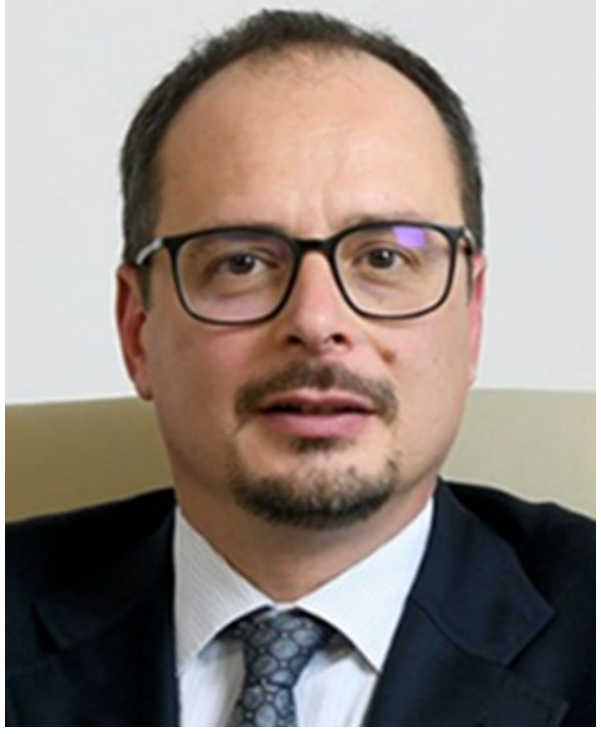

**Rocco Oliveto** is a Full Professor at the University of Molise (Italy) and Director of the Department of Biosciences and Territory. He received his PhD in Computer Science from the University of Salerno in 2008. His research focuses on empirical software engineering, with particular emphasis on machine learning–based recommendation systems, and he has co-authored around 250 publications. He has received multiple ACM SIGSOFT Distinguished Paper Awards and Most Influential Paper Awards. He is also co-founder and former CEO (2017–2025) of Datasound S.r.l., a university spin-off focused on innovative data-driven solutions.

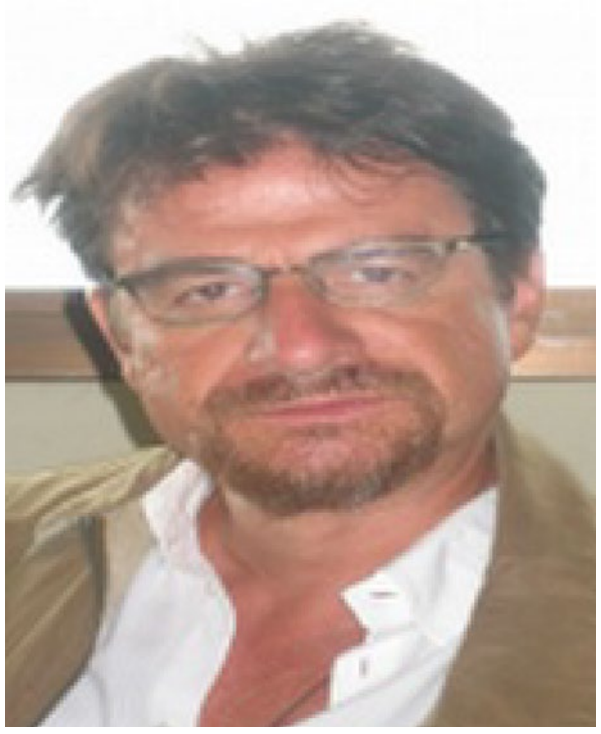

**Remo Pareschi** received his PhD in Artificial Intelligence from the University of Edinburgh. He is an Associate Professor of Computer Science at the University of Molise. He has held research and management positions at the European Computer-Industry Research Centre, Xerox Corporation, and Telecom Italia. His research focuses on software architectures for distributed and intelligent systems, including multi-agent systems, agentic AI, and human–AI interaction. He has been principal investigator on several European-funded research projects and has co-founded start-ups and university spin-offs in machine learning and blockchain technologies. His work combines theoretical foundations with applied system design, with particular attention to the engineering of reliable and adaptive AI-enabled software systems.

## Authors and Affiliations

**Francesco Salzano**[1] · **Cosmo Kevin Antenucci**[1] · **Simone Scalabrino**[1] · **Giovanni Rosa**[1] · **Rocco Oliveto**[1] · **Remo Pareschi**[1]

✉ Francesco Salzano
francesco.salzano@unimol.it

Cosmo Kevin Antenucci
c.antenucci2@studenti.unimol.it

Simone Scalabrino
simone.scalabrino@unimol.it

Giovanni Rosa
giovanni.rosa@unimol.it

Rocco Oliveto
rocco.oliveto@unimol.it

Remo Pareschi
remo.pareschi@unimol.it

1 University of Molise, Campobasso, Italy